\documentclass[11pt,twoside,dvips,a4paper]{article}
\usepackage{graphicx,color}
\usepackage{times}
\usepackage{booktabs}
\usepackage{cite}
\usepackage{epsfig}
\usepackage{feynmp}
\usepackage{color}
\usepackage{a4wide}

\newcommand{\mrm}[1]{\mathrm{#1}}
\newcommand{\tsc}[1]{\textsc{#1}}

\newcommand{\pT}{\ensuremath{p_\perp}}
\newcommand{\TeV}{\,\mbox{Te\kern-0.2exV}}
\newcommand{\GeV}{\,\mbox{Ge\kern-0.2exV}}
\newcommand{\MeV}{\,\mbox{Me\kern-0.2exV}}
\newcommand{\keV}{\,\mbox{ke\kern-0.2exV}}
\newcommand{\eV}{\,\mbox{e\kern-0.2exV}}

\renewcommand{\d}[1]{\ensuremath{\mrm{d}#1}}
\newcommand{\obs}{\ensuremath{\mathcal{O}}}
\newcommand{\sig}{\ensuremath{\sigma}}
\newcommand{\dobs}{\ensuremath{\d{\obs}}}
\newcommand{\dsig}{\ensuremath{\d{\sig}}}
\newcommand{\PS}{\ensuremath{\Phi}}
\def\Re{\mathop{\rm Re}\nolimits}
\def\map{\ensuremath{\kappa}}

\newcommand{\Al}{\tsc{Alpgen}}
\newcommand{\Ar}{\tsc{Ariadne}}
\newcommand{\Co}{\tsc{CompHep}}
\newcommand{\Ca}{\tsc{CalcHep}}
\newcommand{\Fw}{\tsc{Mc@nlo}}
\newcommand{\Hw}{\tsc{Herwig}}

\newcommand{\Mg}{\tsc{MadGraph}}
\newcommand{\Py}{\tsc{Pythia}}
\newcommand{\Sh}{\tsc{Sherpa}}
\newcommand{\Vc}{\tsc{Vincia}}

\begin{document}
\vspace*{-12mm}\begin{minipage}{0.98\textwidth}\footnotesize
\flushright
FERMILAB-PUB-07-160-T\\
Saclay/SPhT--T07/107
\end{minipage}\\[5mm]\hspace*{-0.025\textwidth}
\begin{center}
\Large{\bf A simple shower and matching algorithm}\\[5mm]
\normalsize W.~T.~Giele$^1$, D.~A.~Kosower$^2$, P.~Z.~Skands$^1$\\[5mm]
{\it\footnotesize
\begin{tabular}{rp{15cm}}
1 & Theoretical Physics, Fermi National Accelerator Laboratory, MS106, Box 500, Batavia, IL--60510,
USA\\
2 & Service de Physique Th\'eorique, 
   CEA--Saclay, F--91191 Gif-sur-Yvette cedex, France
\end{tabular}}\\[5mm]
\end{center}

\begin{abstract}
We present a simple formalism for parton-shower Markov chains.  As a
first step towards more complete `uncertainty bands', we incorporate
a comprehensive exploration of the ambiguities inherent in such
calculations. 
To reduce this uncertainty, we then 
introduce a matching formalism which allows a generated event sample
to simultaneously reproduce any infrared safe distribution calculated
at leading or next-to-leading order 
in perturbation theory, up to sub-leading corrections. To enable a more
universal definition of perturbative calculations, we also propose a
more general definition of the hadronization cutoff. Finally, we
present an implementation of some of these ideas for final-state gluon
showers, in a code dubbed \Vc.
\end{abstract}

\section{Introduction}
At present collider energies, the strong (QCD) coupling
strength $\alpha_s$ is sufficiently large that even the most
sophisticated approximations are typically 
reliable only over a limited region of phase space.  
Descriptions which work well for ``hard'' radiation (extra jets, hard
bremsstrahlung) break down rapidly in soft and/or collinear regions
(jet structure, soft bremsstrahlung), and vice versa. In addition, at
scales below a \GeV\ or so, non-perturbative effects must also be
taken into account, and a transition made to a description in terms of
screened charges inside colorless hadrons.  
In this paper, we shall put the
main focus on aspects which are systematically examined in two
complementary perturbative approximations: that of fixed-order
truncations, appropriate for ``hard'' radiation, and that of parton
shower resummations, appropriate for ``soft/collinear'' radiation.

In the past, these two approximations were often pursued independently.
The last decade or so has witnessed rapid progress in our
understanding of how the virtues of each can be used to overcome the
vices of the other, to yield ``matched'' results which attempt
combine the best features of both approaches. 
The earliest concrete approach, due
to Sj\"ostrand and collaborators
\cite{SjostrandEtal}, is
implemented in the \Py\ generator \cite{Sjostrand:2006za} and consists
of re-weighting the first parton shower emission off a hard system
$X$ by a correction factor equal to the ratio needed to reproduce the
tree-level $(X+1)$-parton matrix element. 
This  reweighting relies heavily on
the shower algorithm(s) in \Py\ \cite{Sjostrand:2004ef} 
covering all of phase space
and on the first shower emission being clearly identifiable as the
``hardest''.  A different approach was needed for the coherent 
angular-ordered algorithm \cite{Marchesini:1983bm} used in the \Hw\ Monte
Carlo \cite{Corcella:2000bw}, in which soft gluons can be emitted at
large angles ``before'' harder ones at smaller angles. The approach
developed by Seymour for this purpose
\cite{SeymourMatrixElement}
combines two ingredients: in the
region populated by the shower, emissions are reweighted to produce
the matrix element rate almost as above (generalized to the
angular-ordered case), whereas in the so-called ``dead zone,'' separate
$(X+1)$-parton events are generated according to the appropriate
matrix element, and weighted by a
Sudakov factor.  This technique can be viewed as a precursor to
the modern CKKW approach~\cite{Catani:2001cc}.

Although matching beyond one extra parton was attempted
within the \Py\ framework \cite{PythiaHigherOrder},
the complexity of the problem grows rapidly. 
The CKKW and MLM
\cite{Hoche:2006ph} matching schemes broke through this barrier,
in principle providing a framework for matching
through any number of tree-level matrix
elements, though practical applications are still
limited to including matrix elements for 
at most a handful of additional partons.  
The original CKKW approach is implemented in the \Sh\ generator
\cite{Gleisberg:2003xi}. The  
MLM one is in principle less dependent on the specific implementation,
but is probably most often used with \Al\ \cite{Mangano:2002ea} interfaced
\cite{StandardInterface}
to either \Hw\ or \Py.
The basic idea behind the CKKW scheme 
has since been refined and extended, first via
so-called pseudoshowers introduced by L\"onnblad
\cite{Lonnblad:2001iq,Lavesson:2005xu} in the context of the color dipole
model~\cite{Gustafson:1986db,Gustafson:1987rq} and
implemented in the \Ar\ generator \cite{Lonnblad:1992tz};
and later by Mrenna and Richardson
\cite{Mrenna:2003if} using \Mg\ \cite{Maltoni:2002qb}, 
again interfaced to \Hw\ and \Py.
The most recent advance to be implemented in a widely-used
generator was the subtraction-based loop-level matching
proposed by Frixione and Webber \cite{Frixione:2002ik} that led to 
the \Fw\ \cite{Frixione:2006gn} add-on to the \Hw\ generator.  

\pagestyle{myheadings}
More recently, several groups have presented proposals to 
improve \Fw-style matching~\cite{BetterNLOMatching}; 
to include one-loop contributions in a CKKW-like 
scheme~\cite{NLOCKKW}; to develop a formalism
capable  
of dealing with subleading color and spin effects \cite{Nagy:2007ty};
to include small-$x$ effects 
\cite{SmallX};
and to use 
Soft-Collinear Effective Theory (SCET, see~ref.\cite{Bauer:2001yt}
and references therein) as a framework for matched 
parton showers~\cite{BauerEtAl}.

Making use of a generalization of the matching proposed by Frixione
and Webber, 
our aim is to present a simple formalism for leading-log
leading-color parton showers, constructed explicitly with two main
goals in mind: 1) including systematic uncertainty estimates, and 2)
combining the virtues of CKKW-type matching (matching to tree-level
matrix elements with an
arbitrary number of additional partons) with those of \Fw-type
approaches (matching to one-loop matrix elements). This can be done in
a manner which is simultaneously simple and does 
not introduce any dependence on clustering schemes or $\pT$ cutoffs
beyond those required to regulate explicit subleading logarithms and
hadronization effects. Negative weights will arise in general, in
particular when matching to one-loop matrix elements. 
Within the formalism we present, however, they can
typically be avoided through judicious choice of subtraction terms. 
Moreover, phase space generation only needs to be carried out on
matrix elements which have had their singularities subtracted out, and
hence should be relatively fast. Finally, as a first step towards
making hadronization models (in particular their ``tuning'') less
dependent on the details of the parton shower they are used with, we
propose a generalized definition of the hadronization cutoff.  

As an explicit
proof of concept, we have combined the antenna factorization formalism
\cite{AntennaFactorization} with that of 
dipole showers \cite{Gustafson:1987rq} in
a code dubbed \Vc\ (Virtual Numerical Collider with Interleaved
Antennae) \cite{VinciaPaper}, which is being developed both 
stand-alone and as a plug-in to \Py8\footnote{Many 
thanks to T.~Sj\"ostrand for making this possible.}.

The organization of the paper is as follows: in section
\ref{sec:markov}, we define the general formalism for parton showers,
including a brief discussion of each component. We then
expand the shower into partial cross sections 
of fixed multiplicities of resolved partons.
This expansion is used in section \ref{sec:matching} to
derive a set of matching terms for abitrary tree- and one-loop 
matrix elements, up to corrections of order an infrared
cutoff (hadronization scale) and subleading logarithms. Section
\ref{sec:hadronization} then deals with improving the infrared
factorization between parton showers and hadronization models. The
numerical implementation of many of our ideas, in the form of the \Vc\
code, is then presented in section \ref{sec:vincia}. In section
\ref{sec:nll}, we discuss how one might 
go further in the perturbative expansion. We then round off
with conclusions and outlook in section \ref{sec:conclusion}.

\section{The Shower Chain \label{sec:markov}}
As a starting point, consider a Markov chain algorithm \cite{MarkovChain} 
ordered in some measure of time $t$. Such chains characterize the
development of  
a broad variety of systems. In our application, the system will be 
a set of partons and the r\^ole of $t$ will be played by a measure
of parton resolution scale $Q=1/t$, but
the chain could equally well represent the real-time evolution of a simple
system such as an ensemble of radioactive nuclei.
 Two global quantities characterize such evolution:
 the starting configuration and the duration of the
experiment $t_\mrm{end} - t_\mrm{in}$.  At each differential time
step in between, there is a probability density $A(t)$ for the system
to undergo a non-trivial change. After such a change, 
$A(t)$ itself may change (for example, a nucleus may be replaced
by its decay products). For the
chain to have the Markov property, all that is required is that 
$A(t)$ depend only on the system's present state, not on its
past history. 
This property will turn out to be useful in
the context of higher-order matching in section \ref{sec:matching}.

Let us write the Sudakov factor \cite{Sudakov:1954sw}, the probability
that the system does not change state between two times $t_{\mrm{in}}$
and $t_\mrm{end}$, as:
\begin{equation}
\Delta(t_{\mrm{in}}, t_\mrm{end}) = \exp
\left(-\int_{t_{\mrm{in}}}^{t_\mrm{end}}\hspace*{-3mm} \d{t}\; A(t)\right)~,
\end{equation}
with the understanding that $A$ can depend on the particular system
configuration at time $t_{\mrm{in}}$, and thereby that 
$\Delta$ implicitly has such a dependence as well. 

In the parton shower context, $A(t)$ is the total parton
evolution or splitting probability density. This includes sums
and integrals over all possible types of transitions, such
as gluon-to-gluon pair or gluon-to-quark pair splitting. As our aim is
to resum the leading singularities, $A(t)$ must necessarily be infrared
divergent: there is an infinite probability to radiate a soft or
collinear gluon.  The evolution variable must therefore itself
be infrared safe, such that all the singularities of $A$ correspond
to ``late times'' as defined by $t$ (or, equivalently,
$1/Q$). Specifically, for leading-log evolution $t$ must regulate at
least all leading-log divergences.

Letting $\{p\}_{n}$ denote 
a complete specification of an $n$-parton configuration in the
leading-color limit
(carrying not only information on momenta, but also the color
ordering, flavors, and perhaps polarizations),  
we define the leading-log Sudakov factor by,
\begin{equation}
\Delta(t_{\mrm{n}},t_{\mrm{end}};\{p\}_{n}) =
\exp\Biggl(-\int_{t_{\mrm{n}}}^{t_\mrm{end}}\hspace*{-3mm} \d{t_{n+1}}
\hspace*{-4mm} \sum_{i\in \{n\to n+1\} } \hspace*{-1mm} \int 
\frac{\d\PS^{[i]}_{n+1}}{\d\PS_n} \delta(t_{n+1} -
t^{[i]}(\{p\}_{n+1})) 
A_i(\{p\}_{n}\!\!\to\!\{p\}_{n+1})\Biggr)~,\label{eq:Sudakov}
\end{equation}
where 
$\d\PS_n$ denotes the $n$-particle Lorentz-invariant phase-space
measure, so that $\d\PS_{n+1}^{[i]}/\d\PS_n$ represents the 
branching phase space, and the $t$-ordering is imposed via 
the integral over $t_{n+1}$ together with the $\delta$ function. 
This definition will be the cornerstone
for the remainder of this paper, and so we now devote a few paragraphs
to its explanation. To simplify the notation, we shall usually
let the dependence on $\{p\}_{n}$ be implicit, letting
 $\Delta(t_{\mrm{n}},t_{\mrm{end}}) \equiv
\Delta(t_{\mrm{n}},t_{\mrm{end}};\{p\}_{n})$. 

The first important aspect is that we define the Sudakov factor not for a lone
parton or color dipole, but rather for the $n$-parton configuration as a
whole. The degree to which the evolution of smaller subsystems 
factorize will naturally play an important r\^{o}le but does not
need to be specified explicitly at this point. 
The sum over $i \in
\{n\to n+1 \}$ runs over all possible ways of obtaining $n+1$ partons
from the original $n$ ones. For example, starting with an $n$-parton
configuration of which $n_q$ partons are quarks, an unpolarized
parton shower with four active quark flavors would yield one
term for each quark in the event ($n_q \times q\to qg$)
and five terms for each gluon $(n-n_q) \times ( n_f \times g \to q\bar{q} +
 g\to gg)$. 

The evolution phase space is represented symbolically by the $(n+1)$-parton
phase space for an evolution step of type $i$,  $\d{\PS}^{[i]}_{n+1}$,  
divided by that of the evolving configuration, $\d\PS_n$. 
Its specification requires three variables, along with a mapping
from these variables to the phase space for the emission.
Existing parton-shower
Monte Carlo implementations each choose a different function
for this map. To
name a few known issues, the map may have `dead zones' where it is
zero, and/or it may have regions where several independent emitters
$i$ populate the same point (such double counting is not
necessarily a problem, so long as the sum is properly normalized); 
in some formulations,
the entire event may participate in each branching, in others  
only a specific pair of partons `recoil' off each other; in analytical
leading-log (LL) resummations, a purely collinear map is usually used, 
which slightly violates momentum conservation, and so on. Our point is
not which choice is `best', but that many are possible, each leading
to a different shower evolution. 

In this paper, we shall require that the partons at each step be on
shell and that four-momentum be conserved. 
An $(n+1)$-parton phase space then has $3(n+1)-4$ degrees of freedom:
\begin{equation}
\d\PS_{n+1} = (2\pi)^4 \delta^4\Biggl(\sum_{j=1}^{n+1} p_j - \sum_{i=j}^{n}
\hat{p}_j\Biggr) \prod_{j=1}^{n+1}\frac{\d^3\mathbf{p}_j}{(2\pi)^3
  2E_j}
~,\end{equation}
where we denote the $n$ original momenta
by $\hat{p}$, and the notation for four-vectors is 
$p_i=(E_i,\mathbf{p}_i)$. This represents  
all possible $(n+1)$-parton configurations consistent with energy and
momentum conservation. In the context of evolution, however, we are already
implicitly integrating over all possible $n$-parton configurations,
and we are also explicitly summing over all possible evolution
possibilities $i$ for each such configuration. 
The notation $\d{\PS}^{[i]}_{n+1}/\d\PS_n$ is thus intended to signify the
subdivision of the full phase space into discrete (but possibly
overlapping) regions, each corresponding to a specific  
$n$-parton configuration and a specific $i$. Since the $n$-parton
phase space has $3n-4$ degrees of freedom, this amounts to 
imposing $3n-4$ additional $\delta$ functions and adding the explicit
superscript $[i]$:
\begin{equation}
\frac{\d\PS^{[i]}_{n+1}}{\d\PS_n} = \delta^4\!\left(\sum_{j=1}^{n+1} p_j - \sum_{j=1}^{n}
\hat{p}_j\right) \prod_{j=1}^{n+1}\frac{\d^3\mathbf{p}_j}{(2\pi)^3
  2E_j}\delta^{(3n-4)}\left(
\{\map_i^{-1}(\{p\}_{n+1})\}_{n} - \{\hat{p}\}_n
\right)~,\label{eq:phasespace}
\end{equation}
where $\map_i^{-1}$ is the inverse of the map discussed above.
This inverse can be viewed as a clustering 
definition\footnote{That is, $\map_i^{-1}$ 
``inverts'' the shower in a manner similar to the action, e.g., of~\tsc{Arclus} 
on the \Ar\ shower~\cite{Lonnblad:1992tz} or \tsc{Pyclus} on the $p_\perp$-ordered
\Py\ shower.} 
that, given $i$ and the $(n+1)$-parton
configuration, reconstructs the corresponding ``unevolved'' one,
$\{\hat{p}\}_n$. The requirements on the map $\map_i$ are thus:
\begin{enumerate}
\item For each $i$, a unique inverse of $\map_i$ must exist 
  ($\map_i$ must be injective), 
  however each individual $\map_i$ does not necessarily have to
  cover all of phase space, since we only care about the coverage 
  after summing over $i$ and integrating over  $\d\PS_n$.
\item After summing over $i$ and integrating over $\d\PS_n$, the
  resulting composite map should cover all of phase space (be
  surjective), in order to avoid creating dead zones.
  It does not necessarily have to be one-to-one:
  the $(n+1)$-parton phase space may be covered several
  times so long as this is properly taken into account in the normalization of
  the radiation functions $A_i$ (or more precisely, of their singular parts). 
\end{enumerate}
Obviously, these statements only apply to configurations that are
supposed to be obtainable via shower branchings in the first place, 
and not, for instance, to subleading color topologies such as $Z\to
\bar{q}q gg$ with the two gluons in a color singlet state. 

Below, we shall restrict our attention
to maps corresponding to dipole-antenna showers, such that:
\begin{eqnarray}
\mbox{Dipole Showers:}~~~~~~%
\frac{\d\PS^{[i]}_{n+1}}{\d\PS_n} & = & \frac{\d\PS^{[i]}_{n-2}}{\d\PS_{n-2}}
  \frac{\d\PS^{[i]}_{3}}{\d\PS_{2}}
~~~.\label{eq:dipolefactorization}
\end{eqnarray}
The first factor on the right hand side indicates that we 
we choose $3(n-2)$ $\delta$ functions to express that $n-2$
partons don't move at all. This leaves us with
a dipole-antenna phase space, $\d\PS^{[i]}_{3}/\d\PS_{2}$, 
carrying nine degrees of freedom 
compensated by the six remaining $\delta$ functions. 
Introducing the notation $\hat{a}+\hat{b}\to a+r+b$ which we shall use
for dipole-antenna branchings throughout, 
four of these delta functions embody overall momentum
conservation,  
\begin{equation}
\delta^{(4)}(p_a+p_r+p_b - p_{\hat{a}}
-p_{\hat{b}})~~~. \label{eq:otherpartons} 
\end{equation}
The last two delta functions 
specify the global orientation of the plane spanned by the three
daughter partons in the center of mass of the parent dipole (the
branching plane), relative to the axis of
the parent dipole,  
in terms of two angles, $\theta$ and $\psi$. 
Parity conservation fixes one of them so that the branching plane 
contains the parent dipole axis: 
\begin{equation}
\delta(\theta-\hat{\theta})~~~,
\end{equation}
where $\hat{\theta}$ is the orientation angle of the parent dipole in
a global coordinate system.
The $\delta$ function in $\psi$ fixes the rotation angle of the
daughters around an axis perpendicular to the branching plane. This
angle 
does have a reparametrization ambiguity away from the collinear and soft
limits and hence has the following general form
\begin{equation}
\delta(\psi-\hat{\psi}-\psi_{a\hat{a}})~~~, \label{eq:psi}
\end{equation}
where $\hat\psi$ is the other global orientation angle 
and the reparametrization term $\psi_{a\hat{a}}$ 
will be explored further in section \ref{sec:vincia} (see also
ref.~\cite{theKleissTrick}). For the time being, 
we note only that it must vanish in the soft and collinear limits.

The remaining three integration variables we shall map to
two invariant masses, $s_{ar}=(p_a+p_r)^2$ and $s_{rb}=(p_r+p_b)^2$, 
and the last Euler angle, $\phi$, describing 
rotations of the branching plane around the parent dipole axis. The
antenna phase space then takes the following form:
\begin{eqnarray}
\frac{\d\PS^{[i]}_{3}}{\d\PS_{2}}
& = & \frac{ \lambda\left(s^{[i]},m^2_{\hat a}, m^2_{\hat
  b}\right)^{-\frac12} }{16\pi^2}
\ \d s_{ar} \d s_{rb} 
\frac{\d\phi}{2\pi} \ 
\frac{\d\psi}{\d \hat{\psi}} 
\frac{\d\cos\theta}{\d\cos\hat{\theta}} 
 \\
& = & \frac{1}{16\pi^2 s^{[i]}} \
\d s_{ar} \d s_{rb}
\frac{\d\phi}{2\pi} 
~~~~~~~\mbox{for~~~$m_{\hat{a}}=m_{\hat{b}}=0$~~~;~~~$\psi=\hat{\psi}
  + \psi_{a\hat{a}}$~~~and~~~
  $\theta=\hat{\theta}$}~, \label{eq:masslessphasespace}
\end{eqnarray}
where $\lambda(a,b,c)=a^2+b^2+c^2-2ab-2bc-2ca$ is the K\"all\'en
function, $s^{[i]}$ is the invariant mass squared of the branching dipole, 
and $m_{\hat{a},\hat{b}}$ are the rest masses of the
original endpoint partons. The second line represents the massless case,
with the two orientation angles $\theta$ and $\psi$ fixed as discussed
above.

Immediately following the phase space in eq.~(\ref{eq:Sudakov}) is
a $\delta$ function requiring that the integration variable $t_{n+1}$ 
should be equal to the ordering 
variable $t$ evaluated on the set of $n+1$ partons, 
$\{p\}_{n+1}$, i.e.\ that the configuration after branching
indeed corresponds to a resolution scale of $t_{n+1}$. 
We leave the possibility open that different mappings will be
associated with different functional forms for the post-branching
resolution scale, and retain a superscript on $t^{[i]}$ to denote this.

Finally, there are the evolution or showering
kernels $A_i(\{p\}_{n}\!\!\to\!\{p\}_{n+1})$, representing the
differential probability of branching, which we take 
to have the following form,
\begin{equation}
A_i(\{p\}_{n}\!\!\to\!\{p\}_{n+1}) = 
  4\pi\alpha_s(\mu_R(\{p\}_{n+1})) \ {\cal C}_i  \
  a_{i}(\{p\}_{n}\!\!\to\!\{p\}_{n+1})~~~, \label{eq:A}
\end{equation}
where $4\pi\alpha_s=g_s^2$ is the strong coupling evaluated at
a renormalization scale defined by the function $\mu_R$, 
${\cal C}_i$ is the color factor (e.g.\ ${\cal C}_i=N_c=3$ for
$gg\to ggg$),  and
$a_i$ is a radiation function, giving
a leading-logarithmic approximation to the corresponding
squared evolution amplitude (that is, a parton or dipole-antenna splitting
kernel). When summed over possible overlapping phase-space regions,
the combined result should contain exactly the correct leading soft
and collinear logarithms with no over- or under-counting. Non-logarithmic
(`finite') terms are in constrast arbitrary.
They correspond to moving around inside
the leading-logarithmic uncertainty envelope.  The renormalization scale
$\mu_R$ could in principle be a constant (fixed coupling) or
running. Again, the point here is not to impose a specific choice but
just to ensure that the language is sufficiently general to explore
the ambiguity.

Together, eqs.~(\ref{eq:Sudakov}), (\ref{eq:phasespace}), and (\ref{eq:A}) can be used as a
framework for defining more concrete parton showers. An explicit
evolution algorithm (whether based on partons, dipoles, or other objects)
must specify:
\begin{enumerate}
\item The choice of perturbative evolution variable(s) $t^{[i]}$. 
\item The choice of phase-space mapping
  $\d{\PS}^{[i]}_{n+1}/\d\PS_n$. 
\item The choice of radiation functions $a_i$, as a function of the
  phase-space variables. 
\item The choice of renormalization scale function
  $\mu_R$. 
\item Choices of starting and ending scales. 
\end{enumerate}

The definitions above
are already sufficient to describe how such an algorithm can be
matched to fixed order perturbation theory. 
We shall later present several explicit implementations of these
ideas, in the form of the \tsc{Vincia} code, see section \ref{sec:vincia}.

\def\Branch{\mathcal{P}_{\mrm{branch}}}
Let us begin by seeing what contributions the pure parton shower gives
at each order in perturbation theory. 
Since $\Delta$ is the 
probability of no branching
between two scales, $1-\Delta$ is the integrated 
branching probability $\Branch$.  Its rate of change gives the
instantaneous branching probability over a differential time step 
$\d t_{n+1}$:
\begin{equation}
\begin{array}{rcl}\displaystyle
\frac{\d\Branch(t_n,t_{n+1})}{\d t_{n+1}}
& = & \displaystyle \frac{\d}{\d t_{n+1}}\Big( 1 -
\Delta(t_n,t_{n+1})\Big) \\[4mm]
& = & \displaystyle \sum_i\int 
\frac{\d\PS^{[i]}_{n+1}}{\d\PS_n} \delta(t_{n+1} -
t^{[i]}(\{p\}_{n+1})) 
A_i(\{p\}_{n}\!\!\to\!\{p\}_{n+1})\Delta(t_n,t_{n+1})~.
\end{array} \label{eq:Pbranch}
\end{equation}
This expression 
still contains an explicit
integral over all phase-space variables except $t_{n+1}$. 
The corresponding fully differential distribution of the (time-ordered)
branching probability is simply
the ``na\"ive'' evolution kernel times the Sudakov factor,
\begin{equation}
 A_i(\{p\}_{n}\!\!\to\!\{p\}_{n+1})\Delta(t_n,t_{n+1})
~. \label{eq:diffPbranch}
\end{equation}

We now
seek the complete fixed-order expansion for the distribution of an
observable $\obs$, as computed by the Markov process. 
By definition, $\obs$ is always evaluated on the final 
configuration, reached 
once the Markov chain terminates at $t_\mrm{end}$. 
A convenient physical interpretation of
$t_\mrm{end}$ is as the hadronization
cutoff, $t_\mrm{had}$, beyond which the parton shower is not
evolved; since the evolution would receive $\mathcal{O}(1)$
corrections from hadronization beyond this scale, exclusive event
properties can only be further probed by switching to a
non-perturbative description in that region (alternatively, 
stopping the shower and evaluating $\obs$ on 
  the partons at this scale is still an improvement over fixed-order
  calculations).  
We will return to a generalization of this notion in a later
section; for now merely take $t_\mrm{had}$ as a cutoff in the evolution
variable. This amounts to implicitly treating all radiation (and
hadronization effects) below the
cutoff inclusively, that is summing over additional partons below that
scale. 

Starting from any process, $X$, with differential phase space weight
$w_X$, the parton-shower improved distribution of $\obs$ is: 
\begin{equation}
\frac{\dsig_X}{\dobs}\Big\vert_{\mrm{\color{red}PS}}
 = \int\d{\PS_X}\ w_X\ S(\{p\}_X,\obs)~~~.
\end{equation}
The definition of $w_X$ in the context of matching will be explored in
the next section, but for the pure shower $w_X$ is just the tree-level
matrix element squared for the parent process $X$, possibly subject to 
matrix-element-level cuts or constraints (e.g.\ $Z$ production 
restricted to a window around the $Z$ mass, etc). 
$S$ is a showering operator that generates the Markov chain 
starting from a list of partons $\{p\}_X$. It is defined by:
\begin{equation}\begin{array}{rcl}
\displaystyle \hspace*{-1mm} S(\{p\}_X,\obs) \hspace*{-1mm}
&  \hspace*{-1mm}= \hspace*{-1mm} &\displaystyle \hspace*{-1mm}
  \underbrace{\delta\left(\obs-\obs(\{p\}_{X})\right)
\Delta(t_X,t_{\mrm{had}})}_{\mbox{$X+0$ exclusive above $1/t_\mrm{had}$}}\\[8mm]
& & \hspace*{-0.7cm}+ \displaystyle
\underbrace{\int_{t_{{X}}}^{t_\mrm{had}}\hspace*{-3mm} \d{t_{X+1}}
 \! \sum_i \!
\int 
\frac{\d\PS^{[i]}_{X+1}}{\d\PS_X} \delta(t_{X+1} -
t^{[i]}(\{p\}_{X+1})) \Delta(t_X,t_{X+1})
A_i(...)S(\{p\}_{X+1},\obs)}_{\mbox{$X+1$ inclusive above $1/t_\mrm{had}$}}
~,\label{eq:markov} \hspace*{-1mm}
\end{array}
\end{equation}
where $t_X$, the starting scale for each successive step of 
the evolution, depends implicitly on $\{p\}_X$, the integration over 
$t_{X+1}$ runs over all possible branching scales between $t_X$ and
$t_\mrm{had}$, 
and $A_i(...)$ is defined by eq.~(\ref{eq:A}). 
Expanding the Markov
chain to a few orders will be useful in the context of matching below
and simultaneously illustrates explicitly how the chain works:

\begin{equation}\begin{array}{rcl}
\displaystyle  \hspace*{-7mm}S_X(\{p\}_X,\obs)\hspace*{-1mm}
 & \hspace*{-1mm}= &\nonumber\\[4mm]
&&\hskip -27mm\displaystyle
  \underbrace{\delta\left(\obs-\obs(\{p\}_{X})\right)\Delta(t_X,t_{\mrm{had}})}_{\mbox{$X+0$ exclusive above $1/t_\mrm{had}$}}\\[8mm]
& & \hspace*{-27mm}+ \displaystyle
\int_{t_{{X}}}^{t_\mrm{had}}\hspace*{-3mm} \d{t_{X+1}} \!\sum_i\!\int\!
\frac{\d\PS^{[i]}_{X+1}}{\d\PS_X} \delta(t_{X+1} -
t^{[i]}(\{p\}_{X+1})) \Delta(t_X,t_{X+1})
A_i(...)\nonumber\\[6mm]
&&\hskip -15mm\times\Bigg[\underbrace{
\Delta(t_{X+1},t_{\mrm{had}})\delta\left(\obs-\obs(\{p\}_{X+1})\right)}_{\mbox{$X+1$ exclusive above $1/t_\mrm{had}$}}\\[9mm]
& &  \hspace*{-10mm}+\displaystyle
\int_{t_{X+1}}^{t_\mrm{had}}\hspace*{-3mm} \d{t_{X+2}}  \!\sum_j\!\int\! 
\frac{\d\PS^{[j]}_{X+2}}{\d\PS^{[i]}_{X+1}} \delta(t_{X+2} -
t^{[j]}(\{p\}_{X+2})) \Delta(t_{X+1},t_{X+2})
A_j(...)\nonumber\\[8mm]
&&\times\bigg\{\underbrace{
\Delta(t_{X+2},t_{\mrm{had}})\delta\left(\obs-\obs(\{p\}_{X+2})\right)}_{\mbox{$X+2$ exclusive above $1/t_\mrm{had}$}}
+ \cdots \bigg\}\Bigg]~.%
\label{eq:expansion}
\end{array}
\end{equation}
\noindent Each underbraced term corresponds
to the finite contribution from a specific exclusive final-state
multiplicity (exclusive in the `smeared' sense discussed above, that
is, exclusive above scales $Q_{\mrm{had}}=1/t_{\mrm{had}}$ but inclusive
for smaller scales and hence still infrared safe).
The first underbraced line describes the
shower-improved contribution to the distribution of $\obs$ from `events'
which have no perturbatively resolvable emissions at all, the second 
contributions from events which have exactly one resolved emission, 
and so on.

This expression can now be expanded further, to any fixed order in the
coupling. Each $A$ contains one power of $\alpha_s$, and the
exponentials inside $\Delta$ must also be expanded. The latter expansion
gives rise to higher-order corrections which do not increase the parton
multiplicity, and thus correspond to the `virtual corrections'
generated by the shower.  The explicit $A$ factors, in contrast,
represent the shower approximation of corrections due to
real radiation.

\section{Matching \label{sec:matching}} 

There are several possible definitions of what one might mean by
`matching', reflecting the general concept of making 
two different asymptotic expansions of a given observable agree in an
intermediate region; our first task is thus to establish a
clear nomenclature. In a perturbative calculation, observables
will have expansions in the strong coupling $\alpha_s$.  Each
observable will start at some order $j$ in the coupling, and suffer
corrections at subsequent orders.  As is usual in fixed-order
calculations, we will refer to a calculation of the first order for a
given observable as being of leading order (LO), a calculation accurate to
the following order as next-to-leading order (NLO), and so on.   Because of the
presence of infrared singularities, a given order in an observable
will receive contributions from perturbative amplitudes (matrix
elements) of different loop order. 
We will label perturbative amplitudes of `bare' partons,
complete with their infrared singularities, by loop order\footnote{We will,
however, leave the use of and details of dimensional regularization
and infrared cancellations implicit.}. That is, we reserve the 
nomenclature LO, NLO, etc.\ for \emph{observables} only, and that
of tree-level, one-loop, etc.\ for \emph{matrix elements} only. 

Event samples as produced by a
matched Markov-chain parton shower can be used to measure many
different observables.  We therefore believe it would be misleading to 
characterize them as being of leading or next-to-leading
order. More properly, they should be characterized in terms of which
matrix elements are included in the matching. In this paper, 
for an arbitrary shower initiator process $X$ (the parent process), 
we intend that \emph{``$X$ matched to $X+n$ partons at tree level 
and $X+m$ partons at loop level''} (with $m<n$) 
should fulfill the following:

\def\Ord{{\cal O}}
\begin{itemize}
\item It should resum the leading soft and collinear logarithms to all 
  orders, that is, it should be accurate up to subleading logarithms.
\item For any $j\le n$, it should reproduce the LO distribution of 
  any observable whose expansion starts at order $\alpha_s^j$, 
  up to corrections of order  $\alpha_s^{j+1}$ and/or 
  $Q^2_{\mrm{had}}/Q^2_{\mrm{X}}$, where $Q_{\mrm{had}}$ is a
  hadronization scale and 
  $Q_X$ is a hard scale associated with the $X$ process. 
\item For $k\le m$, it should also reproduce any such distribution
  calculated at NLO, that is up to 
  corrections of $\Ord(\alpha_s^{k+2})$ and $Q^2_{\mrm{had}}/Q^2_{\mrm{X}}$. 
\end{itemize}
The first point
corresponds to the pure parton shower, the second to CKKW- or tree-level
matching, and the third to a generalized variant of loop-level
matching of the sort implemented by \Fw. 
Our purpose here is to combine all three into a unified approach, in which
essentially any tree-level or one-loop matrix element 
could be incorporated with a minimum of effort.

\def\qb{{\overline {\kern-0.7pt q\kern -0.7pt}}}
As an example, consider $Z$ decay. 
A pure parton shower  resums the leading logarithms,
but will only be ``matched'' to $Z+0$ partons at tree level, 
as it will generically
introduce errors of $\Ord(\alpha_s)$ in any observable.  If we match
to $Z+1$ parton at tree level, 
we will correctly reproduce any large-logarithm-free
distribution as predicted using the $Z\rightarrow \qb g q$ matrix
element (the only three-parton leading-order matrix element).  If
we match to $Z+2$ partons at tree level, we will correctly reproduce four-jet 
distributions with only $\Ord(\alpha_s^3)$ corrections, corresponding
to the use of tree-level four-parton matrix elements such as
$Z\rightarrow \qb g g q$. Tree-level matching up to three additional
partons combined with one-loop matching up to two additional partons,
will allow us to reproduce  
four-jet distributions up to corrections of $\Ord(\alpha_s^4)$, in addition to
resumming the leading logarithms, and so forth. 

Let us fix an (arbitrary) observable $\obs$
as representative of the distribution above, and consider
a computation of the 
cross section, or partial width as the case may be, differentially in
$\obs$.  We seek a prescription that will yield
a generated event sample from which distributions can be made that
simultaneously fulfill all the three requirements above.

\def\Matching{{w}}
To specify our matching prescription, we introduce two finite matching
terms at each order in $\alpha_s$, one for resolved radiation (R) and
one for single unresolved and one-loop 
corrections (V). Our expression for the  
matched-shower-improved (MS) distribution is then:
\begin{equation}
\begin{array}{rcl}\displaystyle
\frac{\dsig}{\dobs}\Big\vert_{\mrm{\color{green}MS}}
 & = & \displaystyle\sum_{k=0}^n
\int\d{\PS_{X+k}}\left(\Matching^{(R)}_{X+k} +
 \Matching^{(V)}_{X+k}\right)
\Theta(t_{\mrm{had}} - t(\{p\}_{X+k})
S(\{p\}_{X+k},\obs)~,
\end{array}
\end{equation}
where $\Matching^{(R)}_{X+k}$ is the tree-level matching coefficient
for $X+k$ partons and $\Matching^{(V)}_{X+k}$ is the corresponding 
virtual one. Denoting the couplings present in the parent process
$X$ collectively by $\alpha_X$, these matching terms are of order
$\alpha_X\alpha_s^k$ and $\alpha_X\alpha_s^{k+1}$, respectively. 
One or both of them may be zero in the
absence of matching. The operator $S$ is the same as above, 
embodying an all-orders resummation of both real and virtual
corrections in the leading-logarithmic approximation.  

The tree- and loop-level matching terms $\Matching^{(R)}_{X+k}$ and
$\Matching^{(V)}_{X+k}$ may now be derived by expanding the real and
virtual terms of $S$ separately and, order by order, 
comparing the contribution from 
each fixed parton multiplicity to the observable $\obs$ as
calculated in a fixed-order expansion:  
\begin{equation}
\begin{array}{rcl}\displaystyle
\frac{\dsig}{\dobs}\Big\vert_{\mrm{\color{blue}ME}}
 & = & \displaystyle\hspace*{-3mm}\sum_{k=0}
\int\d{\PS_{X+k}}\;\left\vert\sum_{\ell=0} M^{(\ell)}_{X+k}\right\vert^2
\delta(\obs-\obs(\{p\}_{X+k})~,
\end{array}
\end{equation}
where $k$ still represents the number of legs and 
$\ell$ represents the number of loops.
It goes without saying that matching at incomplete orders will involve
some arbitrariness, which will be explored further below. 

\subsection{``Matching'' to $X+0$ partons at tree level}
Matching to ``$X+0$ partons at tree level'' just means verifying that the
lowest-order expansion of the shower is identical to the lowest-order
parent matrix element, i.e., that all corrections generated by the shower are
of higher order. 
This is trivially true, but let us verify it
explicitly as a first exercise.  
Only the first line of eq.~(\ref{eq:expansion}) is
relevant, with the Sudakov $\Delta$ expanded to unity,
\begin{equation}
\frac{\d{\sig}}{\dobs} \Big\vert_{\mrm{\color{red}PS}}
\sim 
  \int\d{\PS_X}\ \Matching^{(R)}_{X+0} \ 
  \delta\left(\obs-\obs(\{p\}_{X+0})\right)~,
\end{equation}
from which we infer that the trivial condition
$\Matching^{(R)}_{X+0}=|{M}^{(0)}_{X+0}|^2$  ensures 
that the two descriptions match at lowest order.
 
\subsection{Matching to $X+1$ parton at tree level}
At order $\alpha_s$, the parton shower (PS)
generates only one term that contributes to $X+1$ parton
(cf.~the matching definition above): 
\begin{equation}
\begin{array}{lcl}
\mbox{\color{red}PS} & : &  
\displaystyle
\int\d{\PS_X}\;|M^{(0)}_{X+0}|^2
\int_{t_{{X+0}}}^{t_\mrm{had}}\hspace*{-3mm} \d{t_{X+1}} \sum_i\! \int 
\frac{\d\PS^{[i]}_{X+1}}{\d\PS_X}\;\delta(t_{X+1} -
t^{[i]}(\{p\}_{X+1}))A_i(...)\delta\left(\obs-\obs(\{p\}_{X+1})\right)\\
\end{array}~,\label{eq:ps1tree}
\end{equation}
where the preceding (trivial) matching has been used to replace
$\Matching^{(R)}_{X+0}$ by $|M^{(0)}_{X+0}|^2$ and, as above, $i$ runs
over the discrete different branching possibilities, and $A_i$
contains the explicit factor $\alpha_s$.  To match this to the
tree-level $(X+1)$-parton matrix element, we first divide the complete matrix
element phase space into a `resolved' part at early (perturbative)
times $t \le t_{\mrm{had}}$ corresponding to the region populated by
the shower, and an unresolved part $t > t_{\mrm{had}}$ which we will
treat later.
In the region with resolved perturbative radiation, we
will compute the difference in the distribution of $\obs$ by
subtracting the shower term from that of the relevant tree-level 
matrix element (ME),
\begin{equation}
\begin{array}{lcl}
\mbox{\color{blue}ME} & : &
\displaystyle
\int_{t<t_{\mrm{had}}}\hspace*{-4mm}\d{\PS_{X+1}}\;|M^{(0)}_{X+1}|^2
\delta\left(\obs-\obs(\{p\}_{X+1})\right) \\
\end{array}~.\label{eq:me1tree}
\end{equation}
Subtracting the parton shower term, the matching term (MT) at this
partial order, differentially in the additional parton's phase space, is simply
\begin{equation}
\displaystyle
\underbrace{|M^{(0)}_{X+1}|^2}_{\mbox{\color{blue}ME}}
\ \ -  \underbrace{\hspace*{-2mm} \sum_{i\in X\to X+1}\hspace*{-4mm} 
\Theta(t^{[i]}(\{p\}_{X+1})-t_n)
A_i(...)|{M}^{(0)}_{X+0}(\{\hat{p}_i\}_X)|^2
}_{\mbox{\color{red}PS}}~,
\end{equation}
where $i$ now runs over the number of possible contributing 
`parent' configurations for the phase space point in question 
and the $\Theta$ function accounts for the shower not producing
any jets harder than its starting scale (cf.~e.g.\ the discussion of
`power' vs.\ `wimpy' showers in ref.~\cite{Plehn:2005cq}). At such scales,
the shower effectively has a dead zone, and hence 
the matching term becomes just the unsubtracted matrix element. We
show this mostly to illustrate the principle. Because the theta
function only affects the hard emissions, and because the antenna
radiation function captures all the infrared singularities of the matrix
element, the subtracted matrix element is finite in all single-soft
or simple collinear limits. 

For final-state showers, we can start the shower at a
nominally infinite resolution, $Q_H = 
1/t_H \to \infty$, i.e.\ at $t=0$, 
obviating the need for the explicit $\Theta$ function (we defer a
discussion of initial-state showers to future work). We shall
assume this to always be the case and thus define the
equation for the matching term by:
\begin{equation}
\begin{array}{lcl}
\mbox{\color{green}MT} & : &
\displaystyle
\Matching^{(R)}_{X+1}
= |M^{(0)}_{X+1}|^2
\ \ - \hspace*{-2mm} \sum_{i\in X\to X+1}\hspace*{-4mm} 
A_i(...)|{M}^{(0)}_{X+0}(\{\hat{p}_i\}_{X+0})|^2
~.
\label{eq:mt1tree} 
\end{array}
\end{equation}
As before, the hatted momenta $\hat{p}_i$ appearing in ${M}^{(0)}_{X+0}$ 
are a shorthand for the momenta obtained by operating on the
$(X+1)$-parton configuration with an inverse map of type $i$,
\begin{equation}
\{\hat{p}_i\}_X = \{\map_i^{-1}(\{p\}_{X+1})\}_X~.
\end{equation} 

From this discussion, it becomes clear how important it is that the
phase space map allows a relatively clean phase space factorization
such that the nested sums and integrals in eq.~(\ref{eq:ps1tree})
produce a manageable number of subtraction terms with simple borders.
Below, we will construct the \textsc{Vincia} showers explicitly with
this goal in mind.

Before considering the unresolved and virtual corrections, let us
remark on a few noteworthy aspects which appear already at this level.  
We noted above that 
leading-log resummation only fixes the soft/collinear singular terms
of $A$, so that variations in its finite terms are a source of
uncertainty for the shower, and indeed can be used to estimate these
uncertainties.  We see here how
the matching explicitly cancels such variations and hence reduces
the uncertainty:
if $A$ is made ``harder'', then the shower generates more
branchings, but the subtraction term in the matching equation also
becomes larger, making the matching term smaller and compensating the
change. 

An extreme case arises if $A$ is made so large that the
matching term becomes negative in some region of phase space. This
just means that the shower is overpopulating that region relative to
the matrix element, and hence a negative correction is needed to
counter-balance it. The corresponding correction events would have
negative weights, but there is otherwise nothing abnormal about such a
situation. Alternatively, one could switch to a shower re-weighting
procedure as done in \Py\ and thereby maintain positive event weights,
but in the interest of simplicity we shall not consider reweighting
in this paper.

The overall normalization of the parent process under study does change,
by an amount given by the integral
over the matching term.  This includes an integral over
the arbitrary finite terms in $A$.  However, as we have
yet to fix the corresponding virtual term,
this just corresponds to changes within the range of tree-level
uncertainty.

More importantly, this subtraction should in principle be easy to
automate. Given any tree-level matrix element, which these days can
be easily obtained from standard tools like \Co/\Ca
\cite{Pukhov}, \Mg \cite{Maltoni:2002qb}, 
and others, the only additional ingredient needed is a subtraction
term, whose most general form is a sum over 
lower-point matrix elements multiplied by
evolution kernels.  As mentioned above, because the leading singularities
of the resulting subtracted matrix element are absent, it
should be substantially easier to integrate efficiently over phase
space than its unsubtracted counterparts.

Finally, we note that the matching scheme described above 
is inherently incremental. With presently available
methods, a sample of unmatched events cannot easily be modified to
produce a matched sample (except by doing sophisticated
reweightings). Instead, a complete new sample must be generated using
the matched generator. With our method, a pre-generated set of 
events need not be re-generated to improve the matching;
we need only generate an additional set of events
corresponding to the matching term in eq.~(\ref{eq:mt1tree}) and
add it to the first, with the relative weight
of the two samples fixed by the relative integrated cross sections and
the number of events in each sample.  Of course, this only works to the
extent that the particular $A$ chosen for the first
sample is known at the time the second set is generated. The
total cross section corresponding to the combined sample would again be
different than that of the original parton-shower sample, but the
difference is of higher order.

\subsection{Matching to $X+0$ partons at loop level}
To include the full ${\cal O}(\alpha_s)$ corrections to the initiator
process $X$, and thereby fix the normalization of inclusive
$(X+0)$-parton observables (such as the inclusive cross section or the
total width) to NLO, we now turn to the relative order $\alpha_s$ 
corrections to $X$ with zero additional (perturbatively resolved) partons.  
Again, the parton shower only generates one term,  
the first term in eq.~(\ref{eq:expansion}) with the Sudakov
(eqs.~(\ref{eq:Sudakov}) \& (\ref{eq:A})) expanded to order $\alpha_s$:
\begin{equation}
\begin{array}{lcl}
\mbox{\color{red}PS} & : &  
\displaystyle
-\int\d{\PS_X}\;|M^{(0)}_X|^2
\delta\left(\obs-\obs(\{p\}_{X}\right)
\int_{t_{{X}}}^{t_\mrm{had}}\hspace*{-3mm}\d{t_{X+1}}  \!\sum_i\! \int 
\frac{\d\PS^{[i]}_{X+1}}{\d\PS_{X}}
\delta(t_{X+1} - t^{[i]}(\{p\}_{X+1}))A_i(...)\\
\end{array}~~~,\label{eq:ps0loop}
\end{equation}
which, as a consequence of the unitary construction of the shower, 
is essentially identical to the real radiation term 
in eq.~(\ref{eq:ps1tree}).  The only differences are an overall
minus sign, and the fact that 
the observable is here evaluated on the parent configuration ($X$)
rather than on one with an additional emission ($X+1$).

We now wish to find the matching term that, together with the one for
real radiation above, will give the full ${\cal O}(\alpha_s)$
corrections, possibly modulo power corrections in the
non-perturbative cutoff $t_{\mrm{had}}$. To accomplish this, we need
to include two terms from fixed-order matrix elements, one
corresponding to genuine one-loop corrections and another corresponding
to the real radiation below the hadronization cutoff, which was left out above:
\begin{equation}
\begin{array}{lcl}
\mbox{\color{blue}ME} & : &
\displaystyle
\int_{t>t_{\mrm{had}}}\hspace*{-4mm}\d{\PS_{X+1}}|M^{(0)}_{X+1}|^2\delta\left(\obs-\obs(\{p\}_{X+1})\right)
+ 
\int\d{\PS_{X}}2\Re[M_X^{(0)}M_X^{(1)*}]
\delta\left(\obs-\obs(\{p\}_{X})\right)\\ 
\end{array}~.\label{eq:me0loop}
\end{equation}
Let us re-emphasize that the extra parton in the first term is here unresolved
(inclusively summed over) and hence the observable cannot really
depend on it, up to an overall power correction. Within the required
precision, 
the observable dependence is thus the same for all
terms in eqs.~(\ref{eq:ps0loop}) \& (\ref{eq:me0loop}), which we use to justify
lumping them together below. 

The matching term will again be defined by the remainder when
subtracting off the parton shower contribution from the full matrix
element. Differentially in $\d\PS_X$ the matching term becomes:
\begin{equation}
\begin{array}{lcrcl}
\!\!\mbox{\color{green}MT}\hspace*{-3mm}& : &
\displaystyle \Matching^{(V)}_{X} & = &\displaystyle
2\Re[M^{(0)}_XM^{*(1)}_{X}] \  + \  |M^{(0)}_{X}|^2
\sum_i\!\! \int_{\mrm{all}\,t} \!
\frac{\d\PS^{[i]}_{X+1}}{\d\PS_{X}}
A_i(...)
+ \int_{t>t_\mrm{had}}\hspace*{-5.5mm}\d{\PS_{X+1}}\Matching^{(R)}_{X+1}\\
& & & = & \displaystyle
2\Re[M^{(0)}_XM^{*(1)}_{X}] \ + \ |M^{(0)}_{X}|^2
\sum_i\! \int_{\mrm{all}\,t}\!\!
\frac{\d\PS^{[i]}_{X+1}}{\d\PS_{X}}
A_i(...)
\displaystyle \ +  \ \obs(t_X/t_\mrm{had})
~,\end{array}\label{eq:mt0loop}
\end{equation}
where we again used the properties of a clean phase space
factorization and extended the definition of the subtracted
$\Matching_{X+1}$ from eq.~(\ref{eq:mt1tree}) into the unresolved
region.  Because the matched matrix element is free of soft or
collinear singularities, the last term is just a power correction, 
below our required precision. 

Note that a $\Theta$ function restricting the
shower term to contribute only after $t_X$ has again been avoided by
letting the shower populate the entire phase space.
In one-loop matching with additional partons
in the final state, a theta function similar to that in 
eq.~(\ref{eq:mt2tree}) will be present.  We defer a detailed discussion
 to future work.

As usual, the first two terms in eq.~(\ref{eq:mt0loop}) are separately
divergent and a regularization must be introduced before their
(finite) sum can be evaluated. The divergences, which are universal,
are usually regulated using dimensional regularization and the 
cancellation can be performed in a process-independent way. Only the
finite terms must be computed anew for each new process. We thus
believe that this part could also be automated fairly easily,
once the required one-loop matrix elements become available.

We see here how the NLO normalization of inclusive observables 
is fixed. In the matching
of the real radiation term above, the LO normalization changed by the
integral of $\Matching^{(R)}$, a quantity which depends
explicitly on the finite terms in $A_i$. The same variation is
subtracted in eq.~(\ref{eq:ps0loop}).  The final normalization
should accordingly be stable up to higher-order and non-perturbative power
corrections.

\subsection{Matching to $X+2$ partons (and beyond) at tree level}
So far, we have discussed a matching prescription similar to that of
the program \Fw, though we have here attempted to develop a
formalism more readily applicable to the treatment of non-collinear
ambiguities and associated uncertainties. The next step in reducing
these is to include further information from tree-level 
coefficients deeper in the perturbative series.

As mentioned in section \ref{sec:markov},
we shall now limit our attention to evolution
variables which fulfill the Markov property in the strictest sense,
i.e.\ which do not have any explicit memory of the event
history. It then
becomes irrelevant whether a particular $(X+1)$-parton configuration was
obtained by parton showering from $X+0$ partons or from the tree-level
$(X+1)$-parton matching
term. With a uniquely defined ``restart scale'' $t_{X+1}$ in both
cases, the subsequent evolution also becomes the same.

In fact, the Markov property solves nearly the entire problem for us.
We are
interested in the relative order $\alpha_s^2$ double real radiation term ($X+2$
partons) from a shower which we assume has already been matched to
$X+1$ partons above. By virtue of this prior matching, the total
$(X+2)$-parton contribution, for a history-independent evolution
variable, is just given by the parton shower off the tree-level 
$(X+1)$-parton matrix element, here differentially in $(X+1)$-parton
phase space:
\begin{equation}
\begin{array}{lcl}
\mbox{\color{red}PS} & : &  
\displaystyle
|M^{(0)}_{X+1}|^2
\int_{t_{{X+1}}}^{t_\mrm{had}}\hspace*{-3mm} \d{t_{X+2}}  \!\sum_i\!
\frac{\d\PS^{[i]}_{X+2}}{\d\PS_{X+1}} \ \delta(t_{X+2} -
t(\{p\}_{X+2}))A_i(...)\delta\left(\obs-\obs(\{p\}_{X+2})\right)\\
\end{array}~~~,\label{eq:ps2tree}
\end{equation}
which, apart from the replacement $X\to X+1$ and the restriction that
$t$ not depend on $i$ (the Markov property), is identical to the
expression in eq.~(\ref{eq:ps1tree}). The second-level
matching term would then be,
\begin{equation}
\displaystyle
|\Matching^{(R)}_{X+2}|^2
= |M^{(0)}_{X+2}|^2
- \sum_iA_i(...)|M^{(0)}_{X+1}|^2
\Theta(t^{[i]}(\{p_i\}_{X+2})-t_{X+1})~~~,
 \label{eq:naivemt2tree}
\end{equation}
where the $\Theta$ function expresses that the
shower evolution is ordered, i.e.\ that $t_{X+2}$ must come 
after $t_{X+1}$.  That is, the matching coefficient $w$ is obtained
precisely by subtracting the leading singularities as expressed by
the evolution kernel, along with whatever finite terms we have chosen
to include.  It is essentially the same as the subtraction term (up
to the hard-emission modifications due to the theta function) that
would be used for real emission in a next-to-leading order calculation.

Note that the
correction term for configurations which cannot be obtained from a
sequence of two ordered branchings is thus unsubtracted. This is most
obviously the case for subleading color topologies that can arise at
higher orders, like $Z\to qgg\bar{q}$ with the $gg$ pair in a
color singlet state, but in more generality the correction term is
simply the full matrix element for any configuration for which the
parton shower is zero.

Because of the Markov property, this procedure can be repeated for
tree-level matrix elements with an arbitrary number of additional
emissions. However, while the LL antenna functions still only contain 
the leading singularities, the full higher-order matrix
elements will generally contain sub-leading singularities as well. 
This leads to problems with
unwanted contributions coming from matrix elements with ``too many''  
final-state partons. In general, 
all the following terms may appear (after integrating over phase
space), 
\begin{equation}
    \alpha_s^n \prod_{m=0}^{2n} L_{nm} \ln^{2n-m}(Q_1^2/Q_2^2)~, 
\end{equation}
where $Q_{1,2}$ are scales in the problem and $L_{mn}$ are
finite coefficients. For example, at each
order $n$, $m=0$ is the double logarithmic
(eikonal)  term and $m=2n$ is the non-logarithmic (`` finite'') one. We
can now be more specific. Since the shower generates only 
the (leading-color) $m=0$ and $m=1$ pieces exactly, 
the subtraction in eq.~(\ref{eq:naivemt2tree}), beyond $n=1$, may 
leave pieces inside the matching term which would be
divergent were it not for the hadronization cutoff. 
If left alone, this would lead to 
distributions of physical quantities with overly large subleading log
contributions (divergent in the limit the hadronization cutoff is
removed), which is obviously not desirable.  

At $n=2$, corresponding to $(X+2)$-parton matching at tree level, 
these divergences would be removed in an NLL shower (where
``next-to-leading'' here means with respect to the LL shower). Though we do make some 
remarks aimed in this direction at the end of the paper, we note that even if 
we were able to present a complete solution, 
the same problem would then just appear at NNLL level 
when attempting $(X+3)$-parton matching, and so on. To do tree-level
matching beyond one additional parton, clearly, we need a prescription to
consistently regulate the subleading logarithms in tree-level 
$(X+n)$-parton matrix elements. The uncertainty they induce is
within the stated accuracy of the calculation, as they are higher order
in both logarithms and powers of the strong coupling. 

One possibility is to nominally subtract the subleading logarithms as
well in eq.~(\ref{eq:naivemt2tree}), to the extent they are known.
Although the LL shower wouldn't regenerate them, this procedure would
at least cure the problem without affecting the validity of the
approach, up to subleading logarithmic corrections. However, for
tree-level matching to $X$ + many partons, the analytic form of all the
corresponding N$^{\mrm{many}}$LL terms would then have to be
explicitly subtracted, clearly overkill considering
that all we are really after is just a regulator.  

A simpler approach is to place explicit
restrictions on the $\Matching_{X+2}$ phase space, cutting out the
regions where the subleading logarithms become important, for instance
by introducing cuts on parton--parton invariant masses or transverse
momenta. As a rule of thumb one should probably choose the cut to 
be much smaller than the hard scale $Q_X$ 
(so as not to disturb the matching in the hard region)
but still sufficiently large that $\ln(Q^2_X/Q^2_{\mrm{cut}})$ is not
much greater than unity. A back-of-the-envelope estimate would be that 
roughly one order of magnitude between the two scales could be a reasonable 
starting point. 
  
Finally, an alternative approach 
is to use a Sudakov or Sudakov-like function as a
regulating factor. This smoothly suppresses unwanted configurations
while simultaneously maintaining a fixed-order expansion that begins
at unity over all of phase space. 

This gives the
following general form for leading-order matching with any number
of additional partons
\begin{equation}
\begin{array}{lcrcl}\mbox{\color{green}MT} & : & 
\Matching^{(R)}_{X+1+n}
& = & \displaystyle 
\tilde{\Delta}(\{p\}_{X+1+n}) 
\Big(|M^{(0)}_{X+1+n}|^2 \\[3mm] & & & & \displaystyle \hspace*{-1cm}- 
\sum_i A_i(...)|M^{(0)}_{X+n}|^2
\Theta(t^{[i]}(\{p_i\}_{X+n+1})-t_{X+n})\Big)~~;~t<t_{\mrm{had}}~, \label{eq:mt2tree}
\end{array}
\end{equation}
where $\tilde{\Delta}$ is either the Sudakov-like function just
mentioned or, alternatively, just a $\Theta$
function for the cut-off case mentioned above. 

For automated approaches, the $\Theta$ function method
is probably more appropriate for a stand-alone matrix element generator, which
would not have the shower Sudakov readily available, whereas a more
integrated solution could more easily make use of the smoother Sudakov
suppression.  

How does this work in practice?  To generate a sample of events matched
to $n$ additional partons at tree level, we should generate events with zero
through $n$ partons according to probabilities given by the subtracted
matrix elements of eq.~(\ref{eq:mt2tree}), and then evolve each configuration
using the parton shower.

\section{Non-perturbative Corrections \label{sec:hadronization}}
The traditional approach in Monte Carlo parton-shower
generators is to cut the shower evolution off at a low
value of the evolution scale, $Q_{\mrm{had}}$ of ${\cal O}$(1~GeV).  
At this point a transition to a different ``evolution'' is made, 
in the form of 
QCD-inspired phenomenological hadronization models which 
explicitly enforce confinement and other non-perturbative features. 

From the point of view of a perturbative calculation, this cutoff is
simply an arbitrary infrared regulator, below which partons are not
resolved. In the context of the ordered evolution of parton showers,
however, it represents a scale at which non-perturbative components of
the evolution become significant, and hence at which point the
perturbative evolution kernels used in the parton-shower approximation
no longer suffice to describe the physics of events; that is, 
the ``evolution'' should really contain 
large corrections e.g.\ from pion resonances. 

In the context of the \tsc{Vincia} code, we start
by defining the infrared cutoff $Q_\mrm{had}$ in
a more universal way.  Because it simply represents a separation between
regions with ``large'' and ``small'' non-perturbative corrections,
respectively, it is {\it not\/} necessary to tie it to the perturbative
evolution variable. Any infrared-safe phase space contour will
do. For instance, one could easily imagine defining a hadronization
cutoff in terms of dipole-antenna masses applied to a shower which uses
transverse momentum as its evolution variable, as long as the former
regulates all perturbative divergences and the latter
separates off all regions where hadronization corrections are
expected to be large.

We denote the phase space contour defining the
hadronization cutoff for an $n$-parton configuration
by $\Theta_\mrm{had}(\{p\}_{n})$:
\begin{equation}
\Theta_\mrm{had}(\{p\}_{n}) = \Theta(t_\mrm{had} -
t_\mrm{had}(\{p\}_{n}))=\left\{
\begin{array}{cl}
1 & \mbox{in ``perturbative'' region}\\
0 & \mbox{in ``non-perturbative'' region}
\end{array}\right.
\end{equation}
where $t_\mrm{had}=1/Q_\mrm{had}$ is the value of the hadronization
cut-off and its functional form (which may be different from that
of the evolution variable) is given by $t_\mrm{had}(\{p\})$. The
Sudakov factor then takes the form,
\begin{equation}
\Delta(t_n,t_\mrm{end};t_\mrm{had}) = \hspace*{-5mm}\prod_{i\in \{n\to n+1\}
}\hspace*{-5mm} 
\exp\left(-\int_{t_n}^{t_\mrm{end}}\hspace*{-3mm} \d{t_{n+1}} \int 
\frac{\d\PS^{[i]}_{n+1}}{\d\PS_n} \delta(t_{n+1}-t(\{p\}_{n+1}))
\Theta_{\mrm{had}}(\{p\}_{n+1})
A_i(...)\right)~;
\label{eq:hadSudakov}
\end{equation}
for brevity, we have rewritten the sum over $i$ in
eq.~(\ref{eq:Sudakov}) in product form. The perturbative
shower termination scale $t_\mrm{end}$ can now be taken to infinity without
any problem, as the divergences are explicitly regulated by
$\Theta_\mrm{had}$. The probability that the configuration
emits no perturbative (resolved) radiation at all is,
\begin{equation}
\Delta(t_n,\infty;t_\mrm{had}) = \hspace*{-5mm}\prod_{i\in \{n\to n+1\}
}\hspace*{-5mm} 
\exp\left(-\int_{t_n}^{\infty}\hspace*{-3mm} \d{t_{n+1}} \int 
\frac{\d\PS^{[i]}_{n+1}}{\d\PS_n} \delta(t_{n+1}-t(\{p\}_{n+1}))
\Theta_{\mrm{had}}(\{p\}_{n+1})
A_i(...)\right)~,
\label{eq:hadSudakovExcl}
\end{equation}
corresponding to $Q_\mrm{end} = 1/t_\mrm{end}\to 0$.  This probability
is non-vanishing.
The matching equations in section \ref{sec:matching} remain unaltered
by the introduction of this hadronization cut-off, 
except for the replacements $t_\mrm{end}\to\infty$ in the
integral boundaries along with $A\to \Theta_\mrm{had}A$. 

The hadronization cut-off has traditionally been imposed 
in terms of the evolution
variable itself, since, getting one job done well, it usually gets
the other done almost as well. (A few additional
cutoffs are normally imposed, e.g.\ to avoid systems with very low
invariant masses, but those are minor points). 
This has the disadvantage of making the 
region defined to be ``non-perturbative'' 
different from shower model to shower model, and hence a hadronization
model fitted with one shower cannot be used as is with any other shower.

Decoupling the form of the hadronization
cut-off, as proposed here~(\ref{eq:hadSudakovExcl}), 
from the shower parameters (and in
particular the evolution variable), 
would make the non-perturbative modeling more universally applicable. 
This should be true up to the uncertainty inherent in the perturbative
evolution itself. 

This would also be a step towards making it meaningful to compare
different parton showers before hadronization. 
This is in stark contrast
to the present situation, where different parton showers are far from
directly comparable, each having its own cut-off along its own
contour. Fixed-order parton-level calculations
could then be replaced
by parton showers not including hadronization and matched to fixed order
matrix elements as the ``gold standard'' of what is a good 
perturbative QCD calculation.

\section{The \tsc{Vincia} Code \label{sec:vincia}}
We now turn to a proof-of-concept implementation of the ideas
contained in previous sections, in the form of the \Vc\ code
(Virtual Numerical Collider with Interleaved Antennae), implemented
both as a stand-alone program and as a final-state shower plug-in for
\Py8. $H\to gg$, matched to $H\to ggg$ at tree level 
and $H\to gg$ at loop level, 
has also been implemented in both versions,
according to the matching terms defined in section
\ref{sec:matching}. For the plug-in, this includes the 
 possibility of generating negative-weight correction 
events when the shower is overpopulating phase space.

The numerically implemented shower is based on an interleaved
evolution (see e.g.\ ref.~\cite{Sjostrand:2004ef}) 
of systems of color-ordered QCD antennae.  The
implementation discussed here is limited 
to gluons, and uses a strict dipole-antenna factorization
\cite{AntennaFactorization}. (The name `dipole
factorization' is associated with a related NLO formalism due to
Catani and Seymour \cite{CataniSeymour}.) Inserting the massless 
dipole phase space,
eq.~(\ref{eq:masslessphasespace}), into the event Sudakov
eq.~(\ref{eq:hadSudakovExcl}) yields a product of individual
color-ordered dipole Sudakov factors
\begin{equation}
\begin{array}{l}
\displaystyle
\Delta(t_{\mrm{n}},t_{\mrm{end}};\{p\}_{n},t_\mrm{had}) = \\[3mm]
\displaystyle\hspace*{1cm}
\prod_{i}^n
\exp\left(-\int_{t_n}^{\infty}\hspace*{-3mm} \d{t_{n+1}} \int_0^s 
\!\!\d s_{ar}\int_0^{1-s_{ar}} \!\!\!\!\!\!\!\d s_{rb} \int_0^{2\pi}\!\frac{\d\phi}{2\pi}
\delta(t_{n+1}-t(\{p\}_{n+1}))
\Theta_{\mrm{had}}(\{p\}_{n+1})
\frac{A_i(...)}{16\pi^2 s^{[i]}}\right)~,
\end{array}
\label{eq:SudakovDipole}
\end{equation}
where the branching invariants $s_{ar}=(p_a+p_r)^2$ and
$s_{rb}=(p_r+p_b)^2$ are illustrated
in Fig.~\ref{fig:dipolebranching}. 
We now proceed to give explicit forms for each of the
objects required by section \ref{sec:markov} for the definition of
a shower.

\def\ah{\hat{a}}\def\bh{\hat{b}}
\begin{figure}[t]
\begin{center}
\begin{fmffile}{fmfbranch}
\begin{fmfgraph*}(200,100)
\fmfforce{0.1w,0.4h}{lm}
\fmfforce{0.9w,0.4h}{rm}

\fmfforce{0.5w,0.4h}{c}

\fmf{curly,foreground=(0.0,,0.9,,0.0)}{lm,c}
\fmf{curly,foreground=(0.0,,0.9,,0.0)}{c,rm}
\fmf{dashes,right=0.2,foreground=(0.6,,0.6,,0.6),
label=$s=s_{\ah\bh}=s_{arb}$}{rm,lm}
\fmf{dashes,left=0.2,foreground=(0.6,,0.6,,0.6),
label=$s=s_{\ah\bh}=s_{arb}$}{rm,lm}

\fmfv{label=$\bh$}{rm}
\fmfv{label=$\ah$}{lm}

\end{fmfgraph*}\hspace*{1cm}
\begin{fmfgraph*}(200,100)
\fmfforce{0.1w,0.4h}{lm}
\fmfforce{0.9w,0.4h}{rm}

\fmfforce{0.11w,0.24h}{l2}
\fmfforce{0.87w,0.20h}{r2}
\fmfforce{0.52w,0.76h}{t}

\fmfforce{0.5w,0.4h}{c}
\fmf{dots,foreground=(0.0,,0.9,,0.0)}{lm,c}
\fmf{dots,foreground=(0.0,,0.9,,0.0)}{c,rm}

\fmf{curly,foreground=(0.0,,0.0,,1.0)}{l2,c}
\fmf{curly,foreground=(0.0,,0.0,,1.0)}{c,t}
\fmf{curly,foreground=(0.0,,0.0,,1.0)}{c,r2}

\fmf{dashes,right=0.2,foreground=(0.6,,0.6,,0.6),label=$s_{rb}$}{r2,t}
\fmf{dashes,right=0.2,foreground=(0.6,,0.6,,0.6),label=$s_{ar}$}{t,l2}
\fmf{dashes,right=0.2,foreground=(0.6,,0.6,,0.6),label=$s_{ab}$}{l2,r2}

\fmfv{label=$a$}{l2}
\fmfv{label=$b$}{r2}
\fmfv{label=$r$}{t}

\end{fmfgraph*}
\end{fmffile}
\caption{Left: illustration of the two original dipole antennae in a closed
 color-singlet $gg$ system in the center-of-mass. 
 Right: the system after one branching, showing the branching 
phase space invariants $s_{ar}$ and $s_{rb}$. The
 $\phi$ angle corresponds to rotations around the axis of the original
 dipole. The $\psi$ angle corresponds to a rotation of 
 the branching system about an axis perpendicular to the
 branching plane. The ``forbidden angle'' $\theta$, always set to zero in our
 maps below,  
would correspond to rotating the branching plane off axis with respect
 to the original dipole. \label{fig:dipolebranching}}
\end{center}
\end{figure}
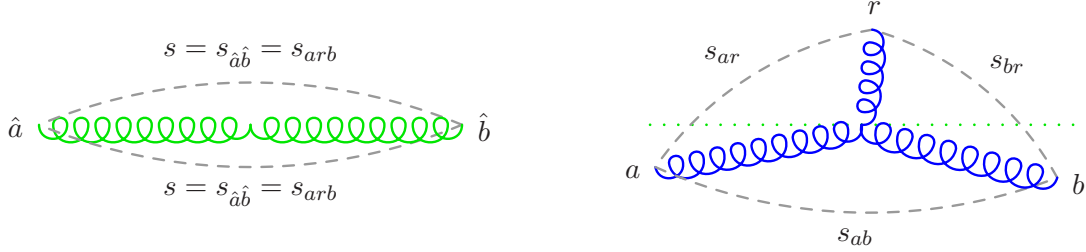

\subsection{Evolution Variable}
\begin{figure}[t]
\begin{center}\vspace*{-5mm}\hspace*{-0.6cm}
\includegraphics*[scale=0.42]{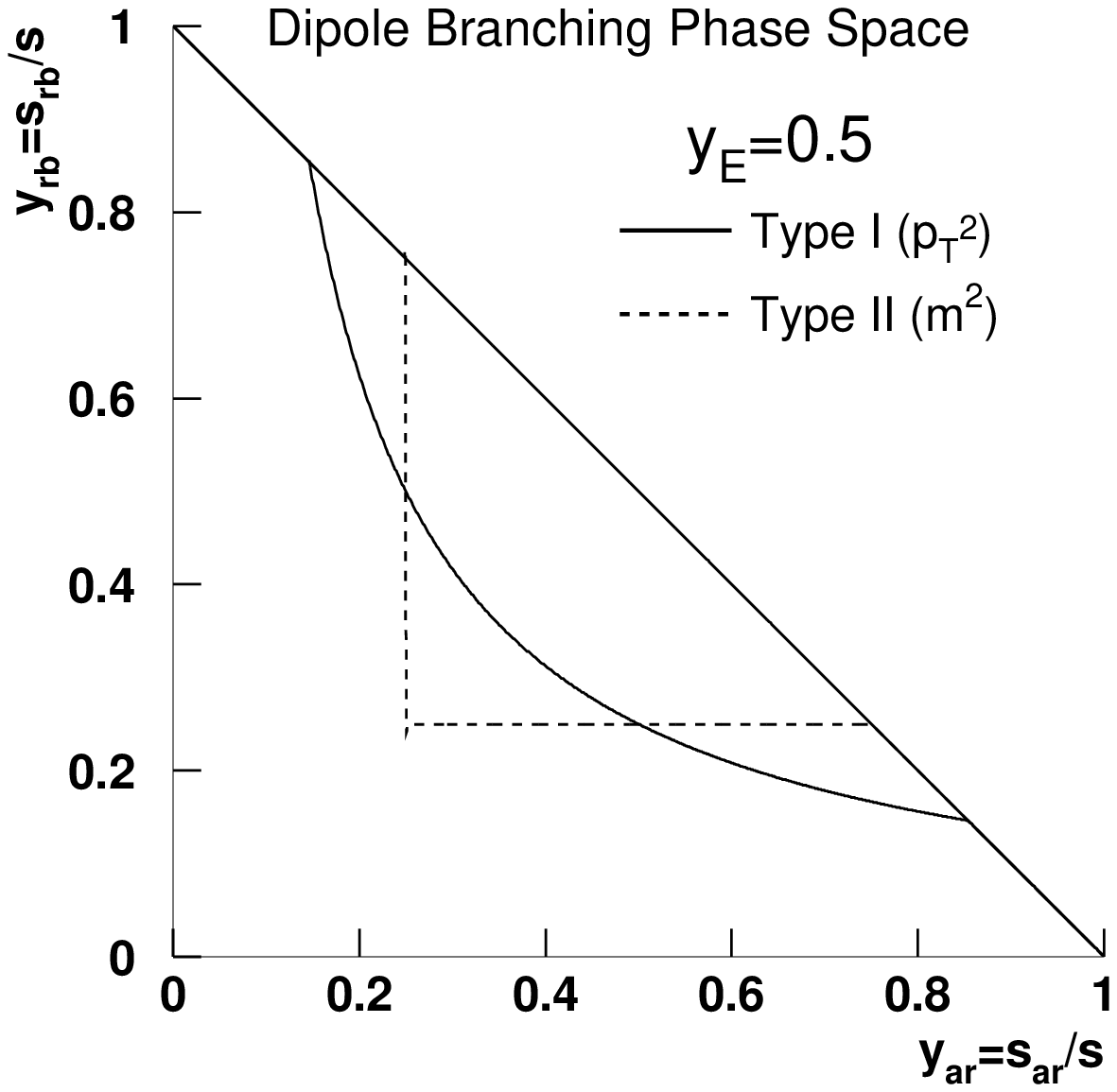}\hspace*{-0.7cm}
\includegraphics*[scale=0.42]{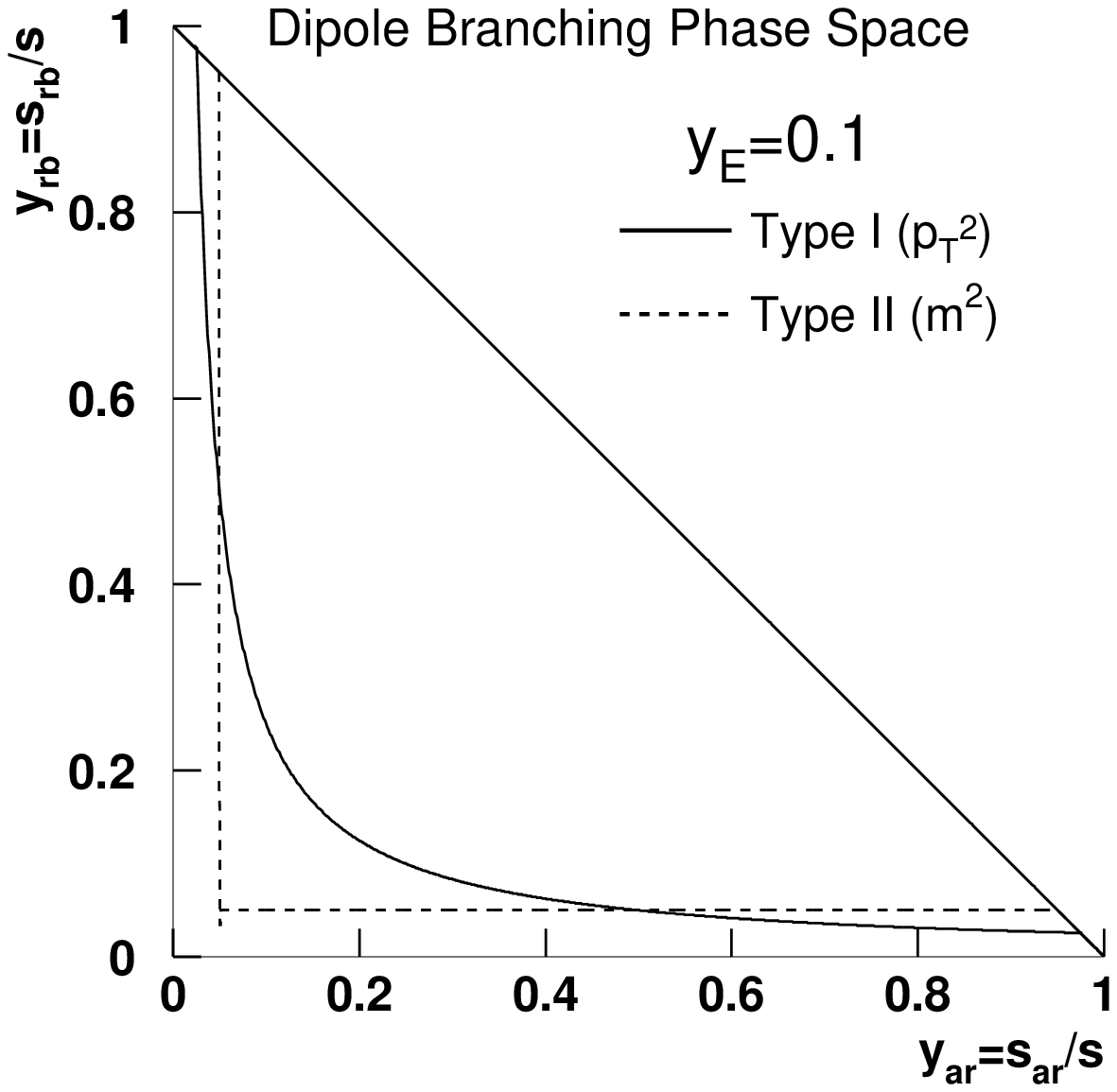}\hspace*{-0.7cm}
\includegraphics*[scale=0.42]{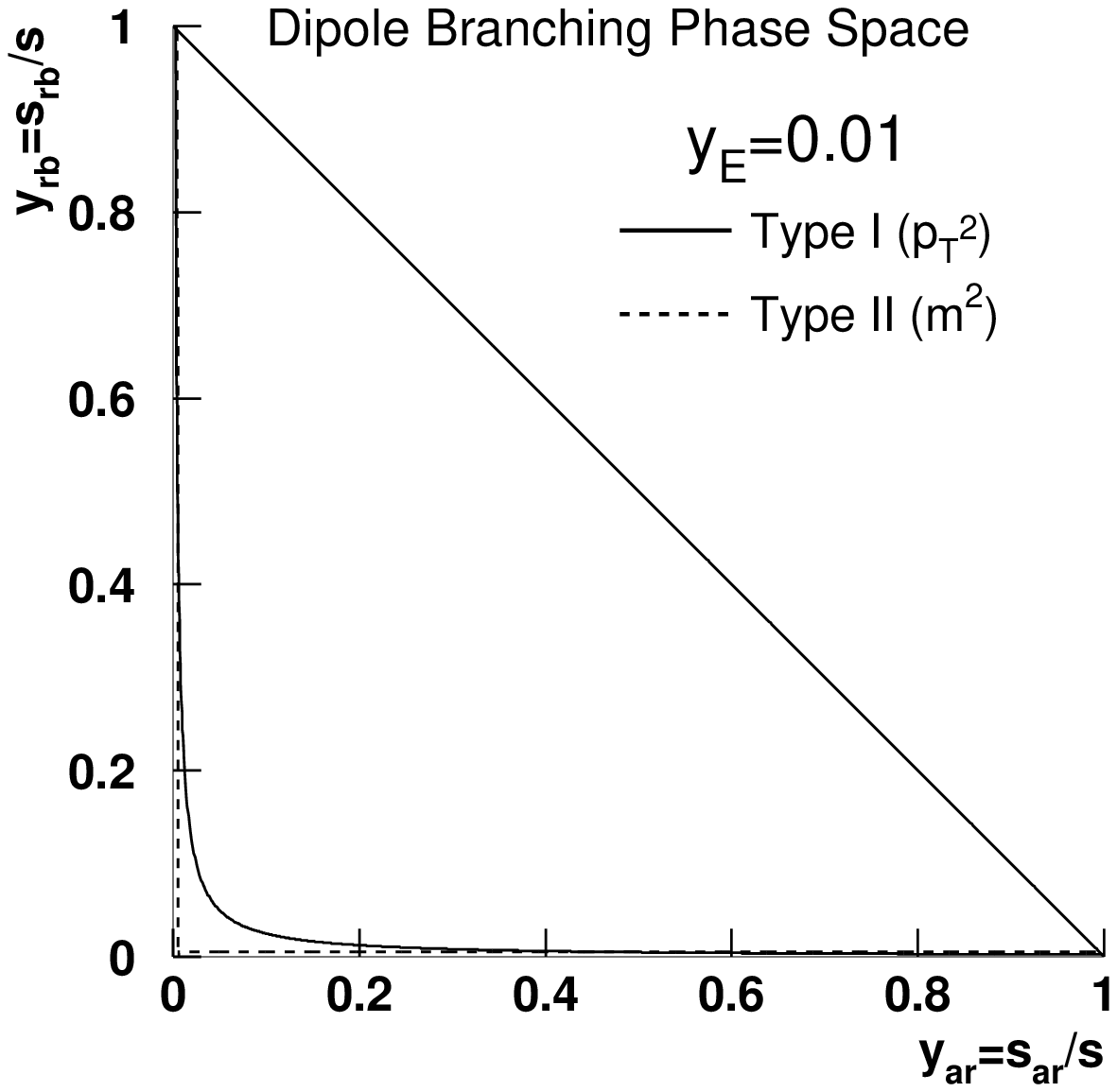}\hspace*{-1cm}\vspace*{-5mm}
\caption{
Illustration of the progression of the evolution variables over phase
space for massless partons: contours corresponding to
$y_\mrm{I,II}=$0.5(left), 0.1(middle), and 0.01(right). On the $x$ and
$y$ axes respectively 
are the rescaled invariants $y_{ar}=s_{ar}/s$ and $y_{rb}=s_{rb}/s$ 
after the branching. The soft singularity is at the lower left corner,
and the collinear singularities lie along the axes.\label{fig:iso-y}}  
\end{center}
\end{figure}

We shall here consider only Lorentz-invariant evolution variables,
$Q_E=1/t$. We have implemented two different choices, corresponding to
ordering in transverse momentum and in dipole mass ($\sim$ parton
virtuality) respectively:
\begin{equation}
Q^2_E = \left\{\begin{array}{rclclcl} 
\mbox{type I}&: & Q^2_{\mrm{I}} 
 & \equiv & \displaystyle 4\frac{s_{ar}s_{rb}}{s} = 4
  p_{\perp\mrm{\Ar}}^2 \\[3mm]
\mbox{type II}&: &Q^2_{\mrm{II}} 
 & \equiv &  \displaystyle2\mrm{min}(s_{ar},s_{rb})
\end{array}\right.~~~,
\label{eq:Q-ordering}
\end{equation}
where the normalizations have been chosen so that the maximum value of
the evolution variable is always the dipole-antenna invariant mass $s$ 
(to avoid cluttering the notation, we now let the superscript $[i]$ be
implicit).  
We will usually work with dimensionless versions of
these invariants, 
\begin{equation}
y_E = \left\{\begin{array}{rclclcl} 
\mbox{type I}&: & y^2_\mrm{I} 
 & = &\displaystyle \frac{Q_\mrm{I}^2}{s} 
 & = & \displaystyle 4\frac{s_{ar}s_{rb}}{s^2} = 4 y_{ar} y_{rb} \\[3mm]
\mbox{type II} & :& y^2_\mrm{II} 
 & = & \displaystyle \frac{Q_\mrm{II}^2}{s} 
 & = & \displaystyle 2\mrm{min}(y_{ar},y_{rb}) 
\end{array}\right.~~~,
\label{eq:y-ordering}
\end{equation}
where $y_i=s_i/s$, so that the maximal value of $y_{\mrm{I},\mrm{II}}$
inside the physical phase space is 
unity. A comparison of iso-$y$ contours for these two variables in the
branching phase space is shown in Fig.~\ref{fig:iso-y}.
Their complementary nature is now readily
apparent. The transverse-momentum or \Ar\ variable
(type I) will categorize a hard but collinear
branching (close to one of the axes) 
as harder than a wide-angle but soft one (close to the origin), whereas the 
the virtuality ordering (type II) will tend to do the opposite. This affects 
which regions act to Sudakov suppress which branchings during the
evolution.

Note, however, that the definitions in eq.~(\ref{eq:y-ordering}) do not
yet completely obey the Markov condition. Because gluons are
indistinguishable, it is not possible to single out 
the radiated parton $r$ 
without knowing the branching history of the configuration. 
In other words, when
showering off a three-gluon configuration with an unspecified
history (e.g.\ from the three-gluon matching term), 
we have several possible choices of what ``restart'' scale to choose, 
depending on which of the partons we decide to call $r$.   
We emphasize that this is not a problem for matching to first order
($X+1$ at tree level and $X+0$ at loop level), since the
history-dependence only has to do with what \emph{restart} scale to
choose and hence, at the earliest, affects the second emission. 
There is no fundamental difficulty in defining variables which
strictly obey the Markov condition, but 
as already discussed in section \ref{sec:matching}, we postpone a 
detailed discussion of this aspect to future work. 

\subsection{Phase Space Map}
We must next choose a phase space map. The restriction to
dipole-antenna phase space factorization,
eqs.~(\ref{eq:dipolefactorization}) \& (\ref{eq:masslessphasespace}), 
already fixes most of the $\delta$ functions: all partons except the two
involved in the splitting are just
``copied'' to the $(n+1)$ configuration. Denoting the branching
antenna pair by $[i]$, eq.~(\ref{eq:otherpartons}) implies
\begin{equation}
p_j = \hat{p}_j ~~~ \forall ~~~ j \notin [i]~.
\end{equation}
The branching antenna pair, denoted $\ah$ and $\bh$, are replaced as
shown in fig.~\ref{fig:dipolebranching}, by a trio of partons,
denoted $a$,
$b$, and $r$.  This replacement conserves energy and momentum,
and keeps all partons at their physical masses.
In the center of mass frame of the parent dipole, the energies are
related to the branching invariants as follows,
\begin{equation}
\begin{array}{rcl}
E_a & = & \frac{s-s_{rb}+m_a^2}{2\sqrt{s}}~,\\
E_b & = & \frac{s-s_{ar}+m_b^2}{2\sqrt{s}}~,\\
E_r & = & \frac{s-s_{ab}+m_r^2}{2\sqrt{s}}~.
\end{array}
\end{equation}
Our discussion here will focus on massless partons, $m_a=m
_b=m_r=0$. 
Since the phase space construction implicitly uses $\delta$ functions
requiring the partons to be on shell, the absolute values of the momenta are
equal to the energies for massless particles. 

\def\mbf#1{{\mathbf #1}}
Staying in the dipole's center-of-mass frame (DCM), 
there is no freedom left to choose the relative angles between the
three daughter partons in the branching plane:
\begin{eqnarray}
 \cos\theta_{ar}&=&\frac{2E_aE_r+m_a^2+m_r^2-s_{ar}}{2|\mbf{p}_a||\mbf{p}_r|}~,
\nonumber\\
 \cos\theta_{ab}&=&\frac{2E_aE_b+m_a^2+m_b^2-s_{ab}}{2|\mbf{p}_a||\mbf{p}_b|}~.
\end{eqnarray}

We still need to fix the orientation of the three daughter partons with
respect to the parent dipole.  This involves three Euler angles.  As noted
in section \ref{sec:markov}, one
of these is fixed by requiring that the branching plane contains the
dipole axis, or equivalently that the normal to the plane be orthogonal to the
dipole axis.  (This imposes parity conservation on the $2\rightarrow 3$
transition.)  Another angle is just the integration variable 
$\phi$ representing
rotations around the dipole axis. While the latter is here chosen
isotropically we note that the matching terms will still (re-)introduce
anisotropies up to the order of the matching.

The remaining ambiguity in the phase space map thus rests entirely
with the last Euler angle, the one corresponding to rotations around an axis
perpendicular to the branching plane. In
the context of \Ar\ \cite{Lonnblad:1992tz}, a choice was made 
which ``least disturbed'' neighbouring
dipoles\footnote{For quark antennae, Kleiss has shown that an optimal
  choice exists \cite{theKleissTrick}, but for gluon antennae the
  situation is less clear.}.
In order to explore the consequences of this ambiguity,
we have so far implemented three discrete possibilities for this
angle, defined by eq.~(\ref{eq:psi}) to be the angle
between parton $a$ and the original parton $\ah$ in the DCM frame, 
\begin{eqnarray}
\psi_{\mbox{\Ar}} & = & \frac{E^2_b}{E_a^2+E_b^2}(\pi-\theta_{ab}) \\
\psi_{\mrm{PS}} & = & \left\{\begin{array}{cl}
0 & ; s_{ar} > s_{rb} \\
\pi-\theta_{ab} & ; s_{ar} < s_{rb} 
\end{array}\right\} \\
\psi_{\mrm{Ant}} & = & 1+\frac{2y_{aa}}{1-y_{rb}}~~~,
\end{eqnarray} 
where $y_{aa}$ in the last line is defined by:
\begin{eqnarray}
f& =& \frac{y_{rb}}{y_{ar}+y_{rb}}~~~,\\
\rho& =& \sqrt{1+4f(1-f)y_{ar}y_{rb}/y_{ab}}~~~,\\
y_{aa}& =&-\frac{(1-\rho)y_{ab}+2fy_{ar}y_{rb}}{2(1-y_{ar})}~~~.
\end{eqnarray}
It is important on physical grounds
that $\psi\rightarrow 0$ smoothly as parton $r$ becomes
collinear with parton $b$ (that is, as $s_{rb}\rightarrow 0$), 
so that parton $a$ becomes aligned with
parent parton $\ah$, and likewise when the roles of $a$ and $b$
and $\ah$ and $\bh$ are simultaneously interchanged.  This ensures that
the daughter system approaches the parent one in this limit.  Otherwise,
however, there are no constraints on $\psi$.  All three alternatives
satisfy this constraint. 

The first choice corresponds to the \Ar\ map, just as our
first evolution variable corresponds to the \Ar\ one
\cite{Lonnblad:1992tz}. We hope this
helps make comparisons between the two approaches simpler.
The second corresponds roughly to conventional parton showers, in
which the non-radiating parton only recoils longitudinally. (Since our
antenna shower does not maintain a clear distinction between which
parton radiates, the one with the largest invariant mass with respect
to $r$ is chosen to play the part of recoiler.)  The last choice is
an example of a more general form; different choices of $f$ could be
used to explore it more fully.

\subsection{Radiation Function}
We have thus far implemented only the $gg\to ggg$ radiation function, 
for which we have used the 
Gehrmann--Gehrmann-De~Ridder--Glover ``global'' 
antenna function$f_3^0$ \cite{GehrmannDeRidder:2005cm}\footnote{
Note that we have changed the non-singular term 
from 2/3 to 8/3, relative to the original paper.},
\begin{equation}
f_3^0(p_a,p_r,p_b) = \frac{1}{s^{[i]}} \Bigg[
(1-y_{ar}-y_{rb})  \bigg(
 \underbrace{\frac{2}{y_{ar}y_{rb}}}_{\mbox{``soft''}} + \underbrace{
 \frac{y_{ar}}{y_{rb}} +
 \frac{y_{rb}}{y_{ar}}}_{\mbox{``collinear''}} 
\bigg) + \frac{8}{3}\Bigg]~.
\end{equation}
In this formula, $s^{[i]}$ is the mass squared of the dipole-antenna. 
``Global'' means that the phase space of each antenna is unrestricted
by overlap with other antennae; the normalization and singularities are 
such that the sum of contributions has the desired
structure. The leading (double logarithmic) singularities correspond
to two invariants vanishing (soft radiation), 
and arise only from one antenna. The
single-logarithmic (collinear) singularities receive contributions from two
neighboring antenna. 

We choose a second-order polynomial in the invariants for the form
of the arbitrary finite terms, imposing only the restriction
that the antenna function be positive definite. Combining $f_3^0$
above with the normalization implied by eq.~(\ref{eq:SudakovDipole}), 
the radiation function for the \Vc\ gluon shower becomes:
\begin{equation}
A(p_a,p_r,p_b) = \frac{4\pi\alpha_s(\mu_R) \ N_c}{s^{[i]}}
\left[(1-y_{ar}-y_{rb}) \left( \frac{2}{y_{ar}y_{rb}} +
 \frac{y_{ar}}{y_{rb}} + \frac{y_{rb}}{y_{ar}} \right) +
 \sum_{\alpha,\beta\ge0}\! C_{\alpha\beta} \ y_{ar}^\alpha
 y_{rb}^\beta\right]~,  \label{eq:Avincia}
\end{equation}
where 
finite terms are parametrized by the
constants $C_{\alpha\beta}$. We can explore systematically
the consequences of making the radiation function harder or softer 
by varying $C_{\alpha\beta}$; e.g., the special case corresponding to 
the $f_3^0$ antenna function can be obtained by choosing $C_{00}=8/3$.
As discussed in section \ref{sec:matching},
matching absorbs these variations in the matching terms, leaving
only the uncertainty due to genuine higher-order terms in the shower.
We can thereby quantify the reduction in the associated uncertainty.

\subsection{Renormalization Scale}
We let the renormalization scale for $\alpha_s$ at each branching be
given either by the evolution scale at the branching, $Q_E^2$, or by the
invariant mass of the dipole being evolved, $s^{[i]}$,
\def\KR{{K_R}}
\begin{equation}
\mu_R = \left\{
\begin{array}{rclcl}
 \mbox{type 1} & : & \mu_1 & = & \KR Q_E \\[2mm]
 \mbox{type 2} & : & \mu_2 & = & \KR \sqrt{s^{[i]}}
\end{array}\right.~~~,
\end{equation}
where $Q_E$ is the evolution variable and we 
allow for an arbitrary prefactor $\KR$ to be applied. 
For example a factor $\KR=1/2$ 
applied to the type I evolution variable would yield a
renormalization scale equal to the \Ar\ definition of transverse
momentum. By default we will use a one-loop running $\alpha_s$, 
but we leave open the option of studying fixed coupling or two-loop
running as well\footnote{In the
\Py8 plug-in, we rely on the $\alpha_s$ implementation in \Py, which
likewise provides these choices.}.
 
\subsection{Starting and Ending Scales} 
For a parent process producing two partons in a decay, such as
$H\to gg$ which we shall consider below, 
we choose the initial starting scale to be
the full phase space, $s$, so that the shower does not have a dead zone. 
After branching at scale $Q_E$, the shower evolution continues from
that scale.   
As already discussed,  
this does imply a slight dependence on the shower history, as
the same configuration can in principle be obtained by different
branchings corresponding to different values of $Q_E$. 
For showers off the three-gluon
matching term, which has no history to provide a unique scale, we
compute the scale corresponding to each possible ordering and select
the smallest of these, as the matching is intended to describe the
hardest emission.  The shower is cut off in the infrared by an
evolution-independent contour, as described in section
\ref{sec:hadronization}. The choices possible for the functional form
of this contour are the same as for the evolution variable,
eq.~(\ref{eq:Q-ordering}).  The history dependence could be eliminated
by using an antenna function restricted to a `wedge' or
`sector' of
phase space; we leave further discussion of this to future work.

\def\DeltaH{{\widehat\Delta}}
\subsection{Shower Implementation\label{ShowerImplementationSection}}
Shower generation proceeds as follows. Given a starting
scale $Q_n$, a trial branching for each antenna dipole is found 
by generating a random number $R\in[0,1]$ and solving for $Q_{n+1}$ in the
following ``trial equation'': 
\begin{eqnarray}
R & = & \DeltaH(Q_n,Q_{n+1})\nonumber \\
& = & \displaystyle 
\exp\Bigg[
-\int_{Q_{n+1}}^{Q_n}\hspace*{-3mm}\! \d{Q_{n+1}} \int_0^{s} 
\!\!\!\d s_{ar}\int_0^{s-s_{ar}}\hspace*{-4mm} \d s_{rb} \
\delta(Q_{n+1}-Q(\{p\}_{n+1})) \frac{\hat{A}(...)}{16\pi^2 s^{[i]}}\Bigg]~,
\label{eq:trial}
\end{eqnarray}
where we use $\DeltaH$ to signify that a nominally larger
branching probability $\hat{A}>A$ may be used to generate these
trials (for instance using an over-estimate of
$\alpha_{s}^{\mrm{max}}>\alpha_s(\mu_R)$, 
no hadronization cutoff, etc); the resulting distribution will then be
corrected by subsequent vetos.

In traditional approaches, an equation for $Q_{n+1}(Q_n,R)$ is
obtained by analytically inverting eq.~(\ref{eq:trial}). 
Since we wish to be able to choose arbitrary evolution variables and
radiation functions, however, we have instead used a more
numerical approach. 

For fixed coupling, the Sudakov factor only depends on one quantity,
the ratio of the evolution scale $Q_{n+1}$ to the starting scale
$Q_n$. Re-expressing the Sudakov factor in terms of dimensionless
ratios of invariants,
\begin{eqnarray}
\hspace*{-5mm}\DeltaH(y_\mrm{trial}) \!\! & \equiv & \!\! \DeltaH(1,y_\mrm{trial})
\nonumber \\
& = & \!\!
\exp\Bigg[-\int_{y_\mrm{trial}}^{1}\hspace*{-3mm}\! \d{y_\mrm{R}} \int_0^1 
\!\!\!\d y_{ar}\int_0^{1-y_{ar}}\hspace*{-5mm} \d y_{rb} \
\delta(y_E-y_E(y_{ar},y_{rb}))
\frac{s^{[i]}\hat{A}(...)}{16\pi^2 }
\Bigg],
\end{eqnarray}
where $y_E=Q_E$ is the dimensionless evolution scale, as defined by
eq.~(\ref{eq:y-ordering}). 
Because the combination
$s^{[i]}\hat{A}(...)$ is independent of $s^{[i]}$,
cf.~eq.~(\ref{eq:Avincia}), this quantity depends
only on a single variable, $y_\mrm{trial}$.   Accordingly,
it is simple to
tabulate it during initialization; we do so using a cubic
spline, performing the integrals inside the exponent either
numerically (via two-dimensional adaptive gaussian quadrature) 
or analytically (as a counter-check).  

We may then solve the equation $R=\DeltaH(y_\mrm{trial})$
numerically for $y_\mrm{trial}$ using the splined version of the
Sudakov and standard root finding techniques.  These
are computationally quite efficient.

The antenna with the largest trial scale is then selected for further
inspection. A $\phi$ angle is chosen uniformly, and 
the remaining degeneracy along the iso-$y$ contour (as 
shown for example in fig.~\ref{fig:iso-y}) 
is lifted by choosing a complementary 
invariant, which we call $z$,  
according to the probability distribution,
\begin{eqnarray}
I_z(y_E,z) & = & \int_0^z \! \d z'
\int_0^1 \!\!\d y_{ar}\int_0^{1-y_{ar}}\hspace*{-4mm} \d y_{rb}
\ \delta(y_E-y_E(y_{ar},y_{rb})) \ 
\delta(z' - z(y_{ar},y_{rb})) 
\frac{s^{[i]}\hat{A}(...)}{16\pi^2 }\nonumber\\[2mm]
& = & \int_{z_\mrm{min}(y_E)}^z \hspace*{-9mm}\d z' |J(y_E,z')|
\frac{s^{[i]}\hat{A}(...)}{16\pi^2 } ~~~, 
\end{eqnarray}
where $|J(y_E,z)|$ is the Jacobian arising from translating
$\{y_{ar},y_{rb}\}$ to $\{y_E,z\}$ and $z_\mrm{min}(y_E)$ is the
smallest value $z$ attains inside the physical phase space 
for a given $y_E$. Since $z$ merely serves as a parametrization of
phase space along an iso-$y_E$ contour, its definition is 
arbitrary, so long as it is linearly independent of $y_E$. Depending 
on the type of evolution variable, we choose $z$ as
\begin{equation}
\begin{array}{lcl}
\mbox{type I} & : & z = y_{rb}~~~,\\[2mm]
\mbox{type II}& : & z = \max(y_{ar},y_{rb})~~~,
\end{array}
\end{equation}
leading to 
the Jacobian factors $|J_\mrm{I}| = 1/(4z)$ and
$|J_\mrm{II}| = 1/2$, respectively, and the phase space boundaries
\begin{equation}
\begin{array}{lclcl}
\mbox{type I} & : & z_\mrm{min}(y_E) = \frac12(1-\sqrt{1-y_E}) & , &
 z_\mrm{max}(y_E) =  \frac12(1+\sqrt{1-y_E})~~~, \\[2mm]
\mbox{type II} & : &  z_\mrm{min}(y_E) = \frac12 y_E& , & 
 z_\mrm{max}(y_E) =  1-\frac12y_E~~~, \\[2mm]
\end{array}
\end{equation}
where the type II case should be divided into two branches, one with
$y_{ar}>y_{rb}$ and one with $y_{rb}>y_{ar}$, each having the phase space
limits given here.

Because the $I_z$ functions depend on two independent variables, $y_E$
and $z$, we have not implemented a splined approach for this task. 
Instead, we use analytical integrals over the two kinds of phase space
regions we are interested in. In generic form, these are
\begin{equation}
I_z(y_E,z) = \frac{\alpha_{s}^{\mrm{max}} N_c}{4\pi} \left[
  S(z)-S(z_\mrm{min}) + K(z)-K(z_\mrm{min}) \right]~,
\end{equation}
with $S(z)$ coming from the soft and collinear singular terms,
\begin{equation}
\begin{array}{lcrcl}
\mbox{type I}  & : &
 S_\mrm{I}(z) & = & \displaystyle\frac{y_E^2}{192z^3} -
  \frac{y_E}{32z^2} + \frac{8+y_E}{16z} -\frac{z}{4} 
   +\frac{-12z + 3z^2 -2z^3 + 12\ln{z}}{6y_E}
~, \\[2mm]
\mbox{type II}  & : &
 S_\mrm{II}(z) & = &\displaystyle
  \frac{1}{24y_E} \left(
   -6(8+y_E^2 +2z -zy_E)z - 8z^3 +3(2-y_E)(8+y_E^2)\ln z\right) 
 ~,
\end{array}
\end{equation}
and $K(z)$ coming from the finite polynomial, 
\begin{equation}
\begin{array}{lcrcl}
\mbox{type I}  & : &
 K_\mrm{I}(z) & = & \displaystyle
  \frac14 \sum_\alpha  \left(\frac{y_E}{4}\right)^\alpha \left[
C_{\alpha\alpha}   
  \ln z
  + \sum_{\beta\ne\alpha}C_{\alpha\beta} 
  \frac{z^{\beta-\alpha}} {\beta-\alpha} \right]
~, \\[5mm]
\mbox{type II}  & : &
 K_\mrm{II}(z) & = &\displaystyle
  \frac12\sum_{\alpha,\beta} C_{\alpha\beta}
 \frac{z^{\beta+1}}{\beta+1}   \left(\frac{y_E}{2}\right)^\alpha
 ~~~\mbox{(
 $C_{\alpha\beta}\leftrightarrow C_{\beta\alpha}$ for $y_{ar}>y_{rb}$)}~,
\end{array}
\end{equation}
with $\alpha,\beta\ge0$. 

With the trial resolution scale $y_E$ and the energy-sharing fraction $z$
now in hand, we can compute $y_{ar}$ and $y_{rb}$.  Together with $\phi$
(chosen above), this gives the complete branching kinematics.  We now
apply a veto, accepting the trial branching with probability,
\begin{equation}
P_\mrm{accept} =
\Theta_\mrm{had}(Q_\mrm{had}(\{p\}_{n+1})-Q_\mrm{had}) 
\frac{\alpha_s(\mu_R(\{p\}_{n+1}))}{\alpha_{s}^{\mrm{max}}}~.
\end{equation}
That is, the branching is only accepted if it is inside the perturbative
region and then only with probability
$\alpha_s/\alpha_{s}^{\mrm{max}}$, which reduces the effective
coupling to the correct value by virtue of the veto algorithm. 
(Note that the event is not thrown away, it
is merely the branching which is vetoed.)
In order to evolve the system further, we repeat the steps above.
The trial branching scale becomes the new starting scale, whether the
above branching was vetoed or not.

The evolution continues until 
there is no perturbative evolution space left (the equivalent
of reaching the hadronization cutoff in our terminology). In the
current implementation, we consider this condition satisfied for a 
given antenna if ten consecutive trials are rejected due to the
$\Theta_\mrm{had}$ condition. 

\subsection{Matching Implementation}
In this paper, we restrict ourselves to matching at first order (at tree
and loop level) 
for a scalar decaying into two gluons via an effective point
coupling. By first order matching we mean that, in addition to 
$\Matching^{R}_2=|M^{(0)}_2|^2$, we include
the matching coefficients $\Matching^{(R)}_3$ and
$\Matching^{(V)}_2$. For the decay process $H\to gg$ the
subtracted matrix elements are relatively easy to obtain.
Given the Born squared matrix element $|M^{(0)}_2|^2$ for $H\rightarrow gg$
we find for the 2-gluon matching term at one loop
\begin{equation}
\Matching^{(V)}_2=
\left(1+\left(\frac{\alpha_s(\mu_R)N_c}{2\pi}\right)
\left[\frac{87}{6}+\frac{11}{3}\log\left(\frac{\mu_R^2}{M_H^2}\right)+\sum_{\alpha,\beta}C_{\alpha\beta}
\frac{\alpha!\beta!}{(2+\alpha+\beta)!}
\right]\right)|M_2^{(0)}|^2
\end{equation}
and for the 3-gluon matching term
\begin{eqnarray}
\Matching^{(R)}_3&=&\frac{8\pi\alpha_s(\mu_R)N_c}{M_H^2}\Big(8-F_{123}-F_{231}-F_{312}\Big)|M_2^{(0)}|^2\nonumber \\
&=&\frac{8\pi\alpha_s(\mu_R)N_c}{M_H^2}
\Big(8-3C_{00}-(C_{10}+C_{01})-C_{11}\left(y_{12}y_{23}+y_{23}y_{31}+y_{31}y_{12}\right)+\cdots\Big) |M_2^{(0)}|^2~,\nonumber \\
\end{eqnarray}
where
\begin{equation}
F_{arb}=\sum_{\alpha,\beta\ge 0} C_{\alpha\beta}y_{ar}^\alpha y_{rb}^\beta~.
\end{equation} 
We note that by taking $C_{00}=8/3$ and all other coefficients equal to
zero, the three-gluon matching term is zero. 
This means that the complete $H\to ggg$ amplitude has been absorbed into the
Sudakov factor. 

\subsection{Preliminary Results}
We now turn to a comparison between results obtained using a few
different parameter and variable choices with the \Vc\ code (using the
\Vc\ plug-in module with \Py8.086). Recall that we
are here studying pure gluon evolution in the fictitious decay of a
scalar to two gluons. We thus intend these results mostly for
illustration of the method. 
We use the type I evolution variable with a one-loop running
$\alpha_s$. The hadronization scale
is chosen to be $Q_\mrm{I} = 1 \GeV$. 

\begin{figure}[t]
\begin{center}
\vspace*{-5mm}
\begin{tabular}{c}
\hspace*{-1cm}
\includegraphics*[scale=0.65]{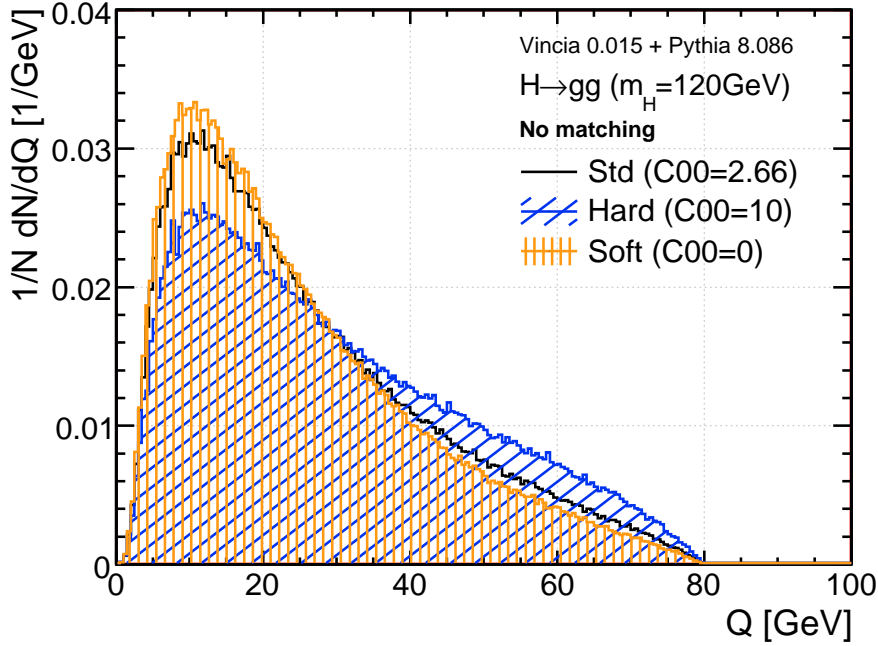}\hspace*{-1cm}\vspace*{-5mm} 
\end{tabular}
\caption{Type I evolution scale for the first shower
  branching, 
symmetrized over $a$, $r$, and $b$: 
  $Q=\min(Q_\mrm{I}(a,r,b),Q_\mrm{I}(r,b,a),
  Q_\mrm{I}(b,a,r))\sim 2\pT$, for $H\to gg$ with $m_H=120\GeV$, 
showered with three different choices of the finite terms in the \Vc\
shower, corresponding to ``standard'' ($C_{00}=2.66$),
``hard'' ($C_{00}=10$), 
  and ``soft'' ($C_{00}=0$)
  variations. 
\label{fig:1stbranching}}
\end{center}
\end{figure}

The plot in fig.~\ref{fig:1stbranching} 
illustrates the distribution of the 
(symmetrized) type-I resolution scale for three-parton configurations, 
obtained
with unmatched \Vc\ for ``soft'' (all $C_{\alpha\beta}=0$), ``standard''
($C_{00}=2.66\sim8/3$) and ``hard'' ($C_{00}=10$) variants. 
For all curves, $\mu_R=Q_\mrm{I}/2\sim\pT$. The point $Q=80\GeV$ 
corresponds to the ``Mercedes'' configuration. While the variations
greatly affect the shape of the distribution, 
the peak position remains fairly stable, here at around a tenth of the
original mass. 

\begin{figure}[tp]
\begin{center}\vspace*{-5mm}\small
\begin{tabular}{cc}
\includegraphics*[scale=0.3]{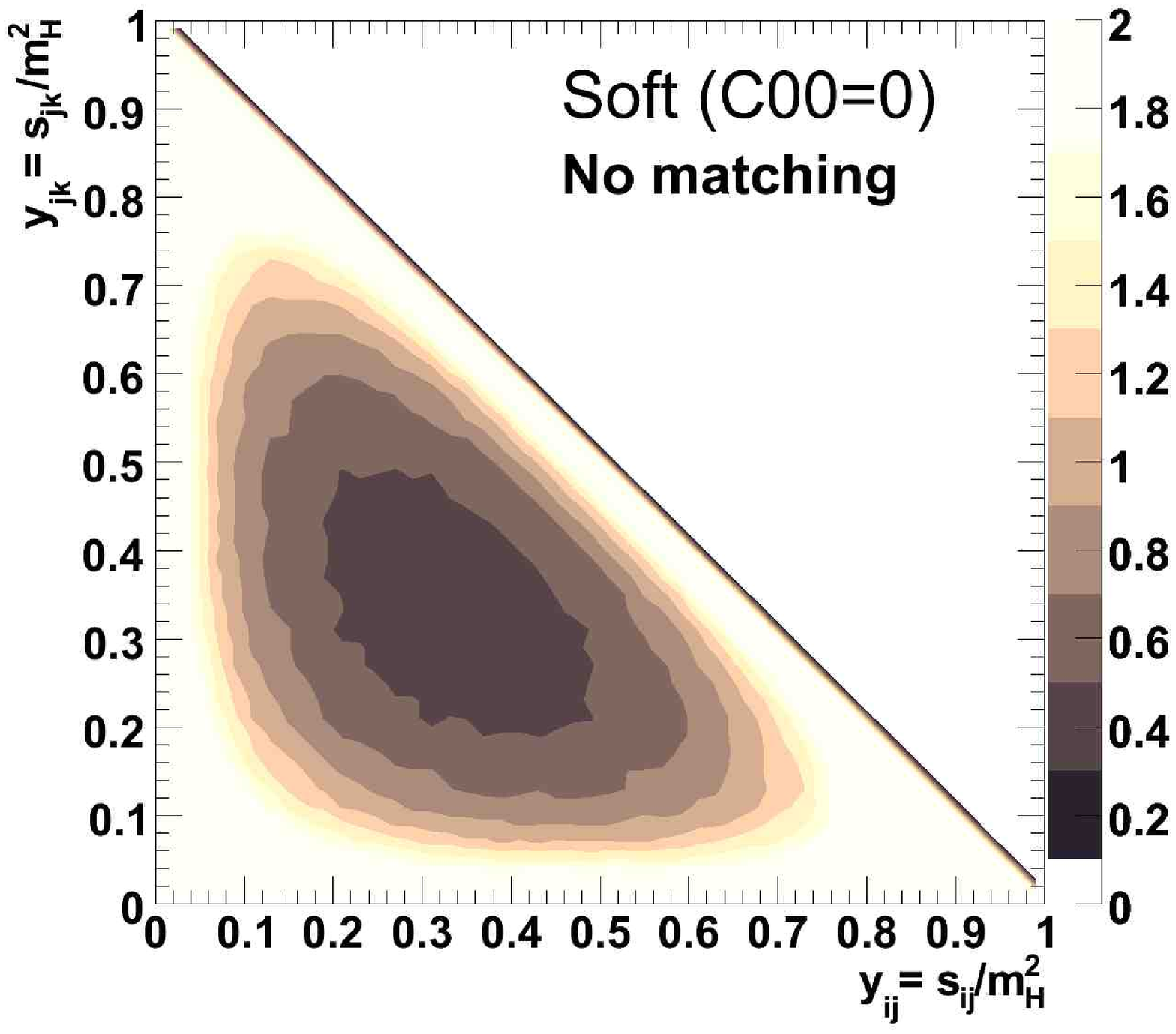} &
{\includegraphics*[scale=0.32]{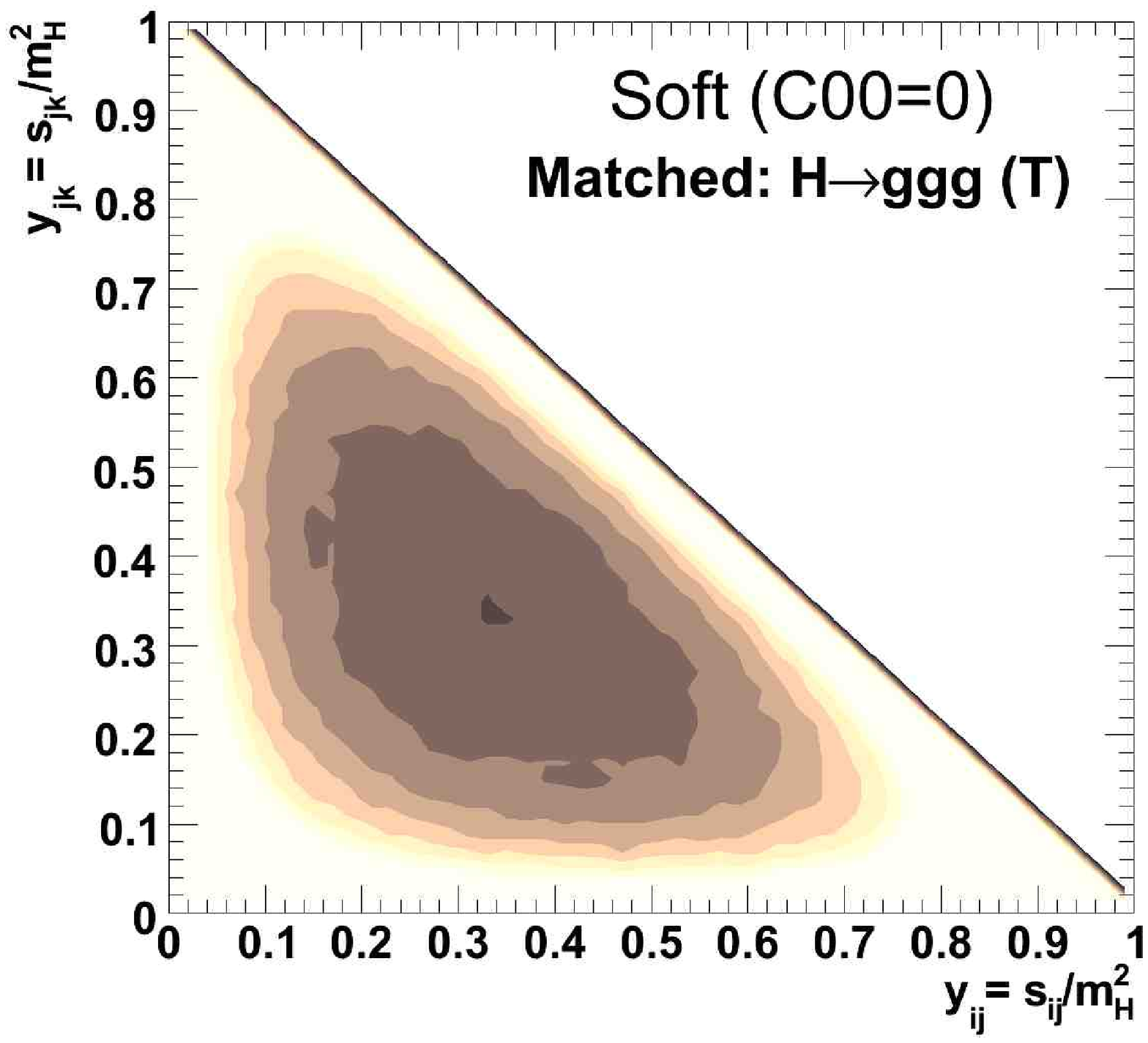}}
\\[-8.6mm]
\includegraphics*[scale=0.3]{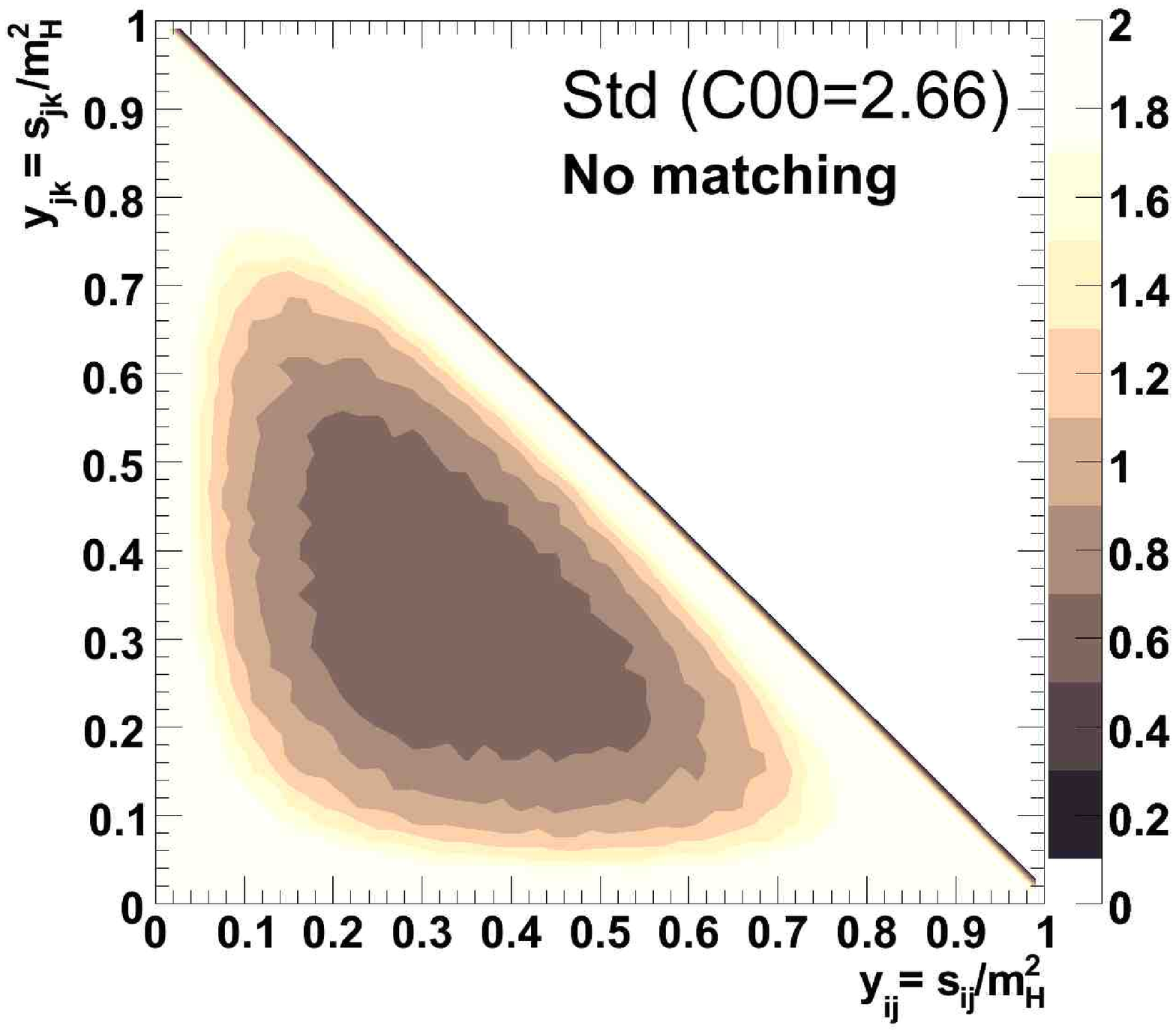}& 
\\[-8.6mm]
\includegraphics*[scale=0.3]{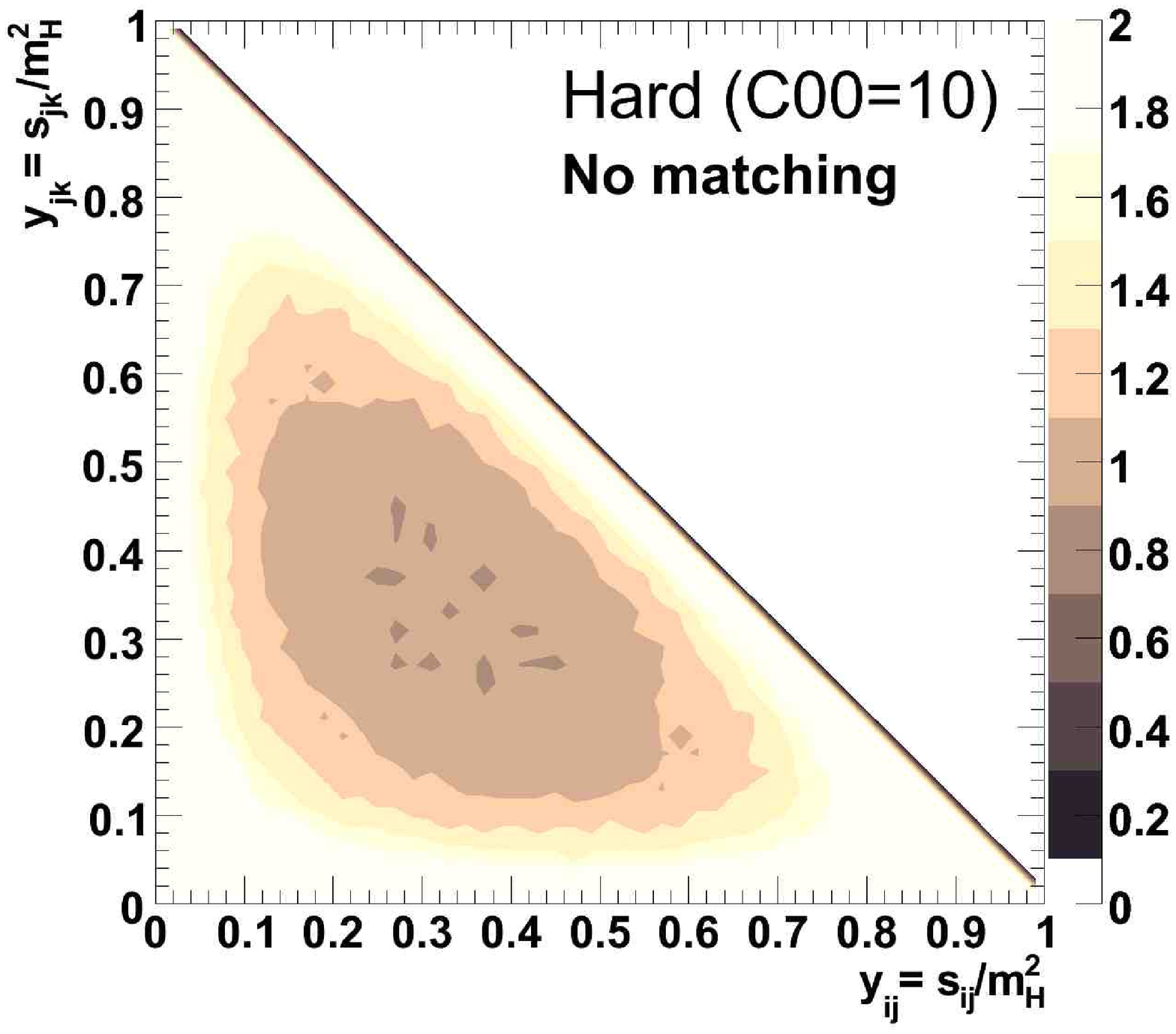}
&
{\includegraphics*[scale=0.32]{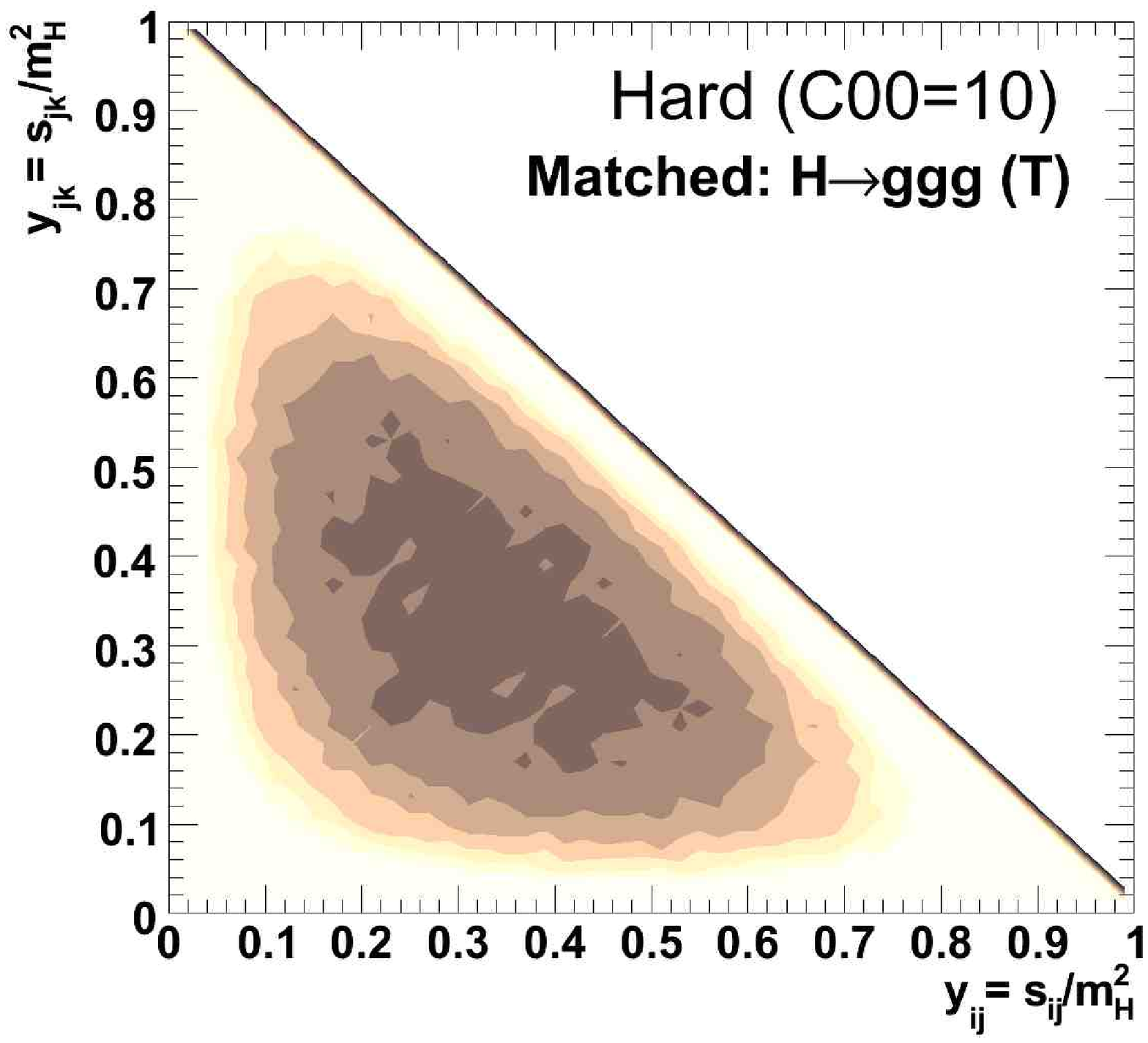}}
\end{tabular}\vspace*{-6mm}
\caption{Phase space population $1/N dN/dy_{ij}/dy_{jk}$, for
 three-gluon configurations,
 symmetrized over all combinations of $i$, $j$, and $k$.
Top left: the ``soft'' 
\Vc\ shower off two-parton configurations, with all $C_{ij}=0$. Top right:
as in top left,
 but including the tree-level $H\to ggg$ matching term. 
Note how the radiation
``hole'' in the center is filled in slightly. 
Middle: the ``standard'' \Vc\ shower, which absorbs the tree-level 
$H\to ggg$ matrix element correction. 
Lower left: same as
upper left, but with $C_{00}=10$, corresponding to a ``hard'' \Vc\ 
shower. Lower right: as in the lower left, but including tree-level
 matching. In this case, the  $H\to ggg$ correction
 is negative, reducing the population of the central
 region.
\label{fig:phasespace}}
\end{center}
\end{figure}

To investigate how matching  reduces this uncertainty, 
fig.~\ref{fig:phasespace} shows the two-dimensional 
phase space population for three-parton configurations corresponding to five
different settings of the \Vc\ plug-in, from top left to bottom right:
soft (unmatched), soft (matched), standard, hard (unmatched), hard
(matched), where matched here refers to matching to $H\to ggg$ at tree
level. As in fig.~\ref{fig:iso-y}, the dipole phase space
is represented as a triangle in the two phase space invariants,
$y_{ij}=s_{ij}/s$ and $y_{jk}=s_{jk}/s$, here 
symmetrized over $i$, $j$, and $k$ (because gluons are
indistinguishable). The dark
color in the center of the plots indicates low probability, with warmer
colors (lighter shades) 
denoting increasing probabilities towards the corners and
sides. In order to focus on the hard central region, the color scale has
been forced to white for $1/NdN/dy_{ij}dy_{jk}\ge2$, thus
``whiting out'' the strong peaking towards the corners and sides of the
triangle which would otherwise dominate.  
In the top row, corresponding to the soft shower, 
the matching fills in missing configurations in the near-``Mercedes'' region.
In the middle row, no matched plot is shown, because
setting $C_{00}=8/3$ corresponds to exponentiating the
matrix element itself and hence the shower already produces the
``correct'' result in this case. For the lower two plots, the shower is
significantly harder than the matrix element. In this case, the code
responds by generating a matching term with negative
weight, effectively reducing the population of the hard region when
``added'' to the unmatched events. 
The reduction in uncertainty for this observable 
is evident by comparing the variation from top to bottom
on the left, with the variation on the right. (The somewhat
odd-looking contours are merely an artifact of \tsc{Root}'s 
contour algorithm operating on binned histograms.)

\begin{figure}[t]
\begin{center}
\vspace*{-5mm}\begin{tabular}{cc}
\hspace*{-6mm}
\includegraphics*[scale=0.43]{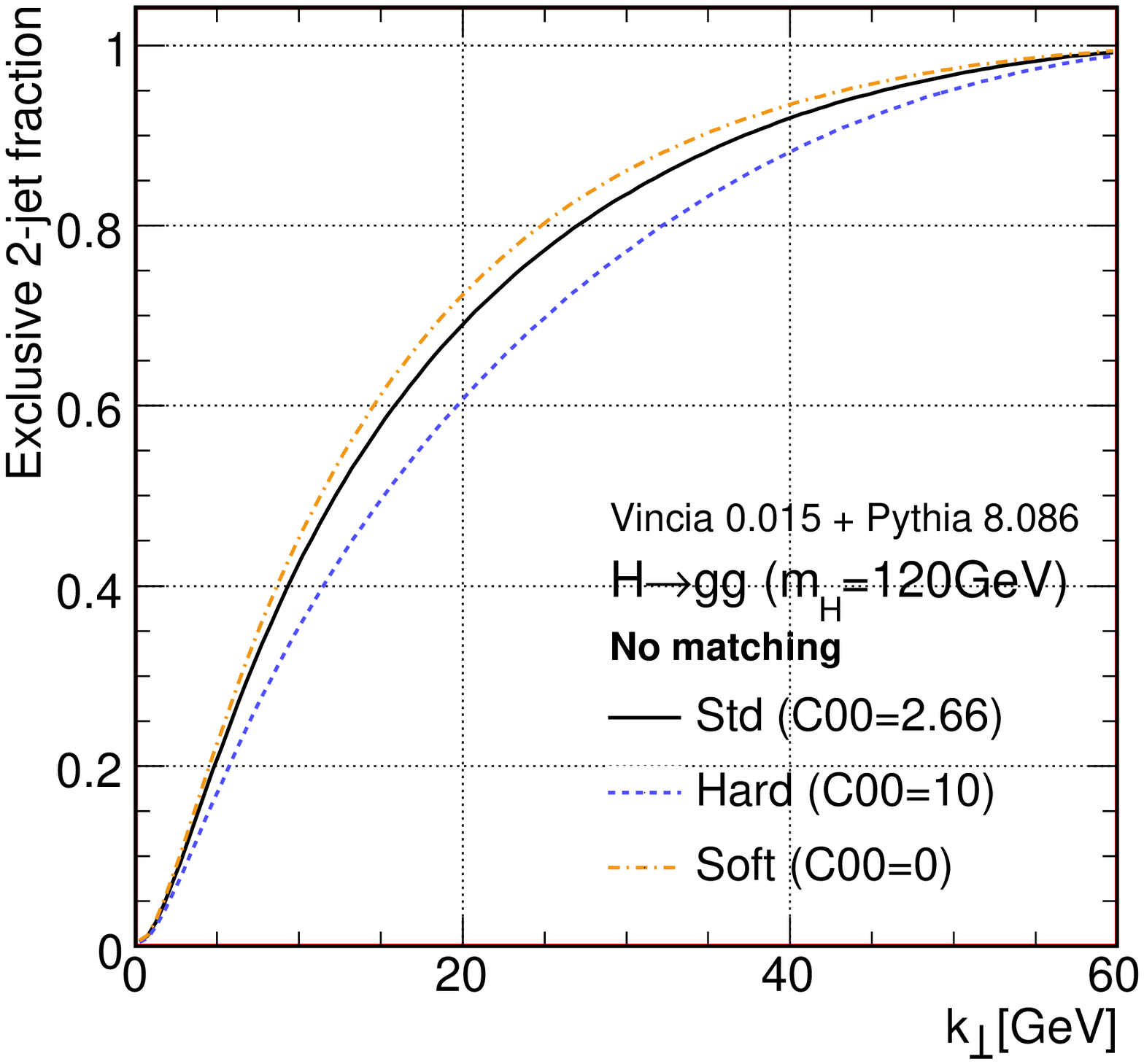}\hspace*{-5mm}&\hspace*{-5mm}
\includegraphics*[scale=0.43]{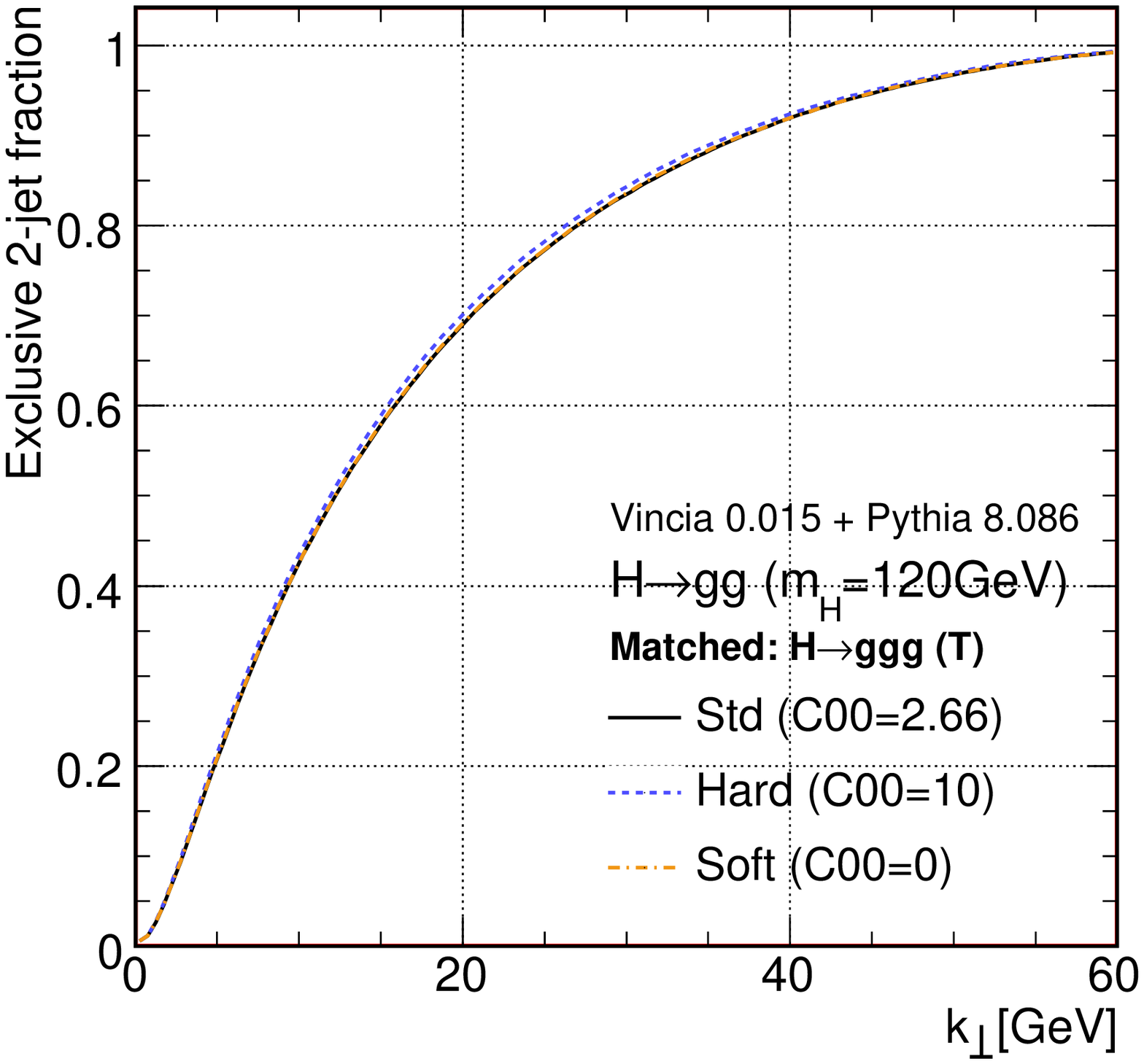}\hspace*{-1cm}\vspace*{-7mm}
\end{tabular}
\caption{The number of exclusive two-jet configurations as a function
  of \tsc{Durham} $k_\mrm{\perp}$. 
Labels are identical to those in fig.~\ref{fig:1stbranching}, except
  that we have here used a different renormalization scale choice, 
$\mu_R=Q_\mrm{II}/2$, translating to $\mu_R= m_H/2$ for the first
  branching over all of phase space. 
\label{fig:no2j}}
\end{center}
\end{figure}
In fig.~\ref{fig:no2j} we show the number of two-jet
configurations as a function of the type-I resolution scale, roughly
equivalent to the Sudakov factor expressed in this variable, before (left)
and after (right) matching to the tree-level $H\to ggg$ matrix
element. As could be expected, the uncertainty on this observable is
greatly reduced by matching to this level. 

\begin{figure}[t]
\begin{center}
\vspace*{-5mm}\begin{tabular}{cc}
\hspace*{-6mm}
\includegraphics*[scale=0.43]{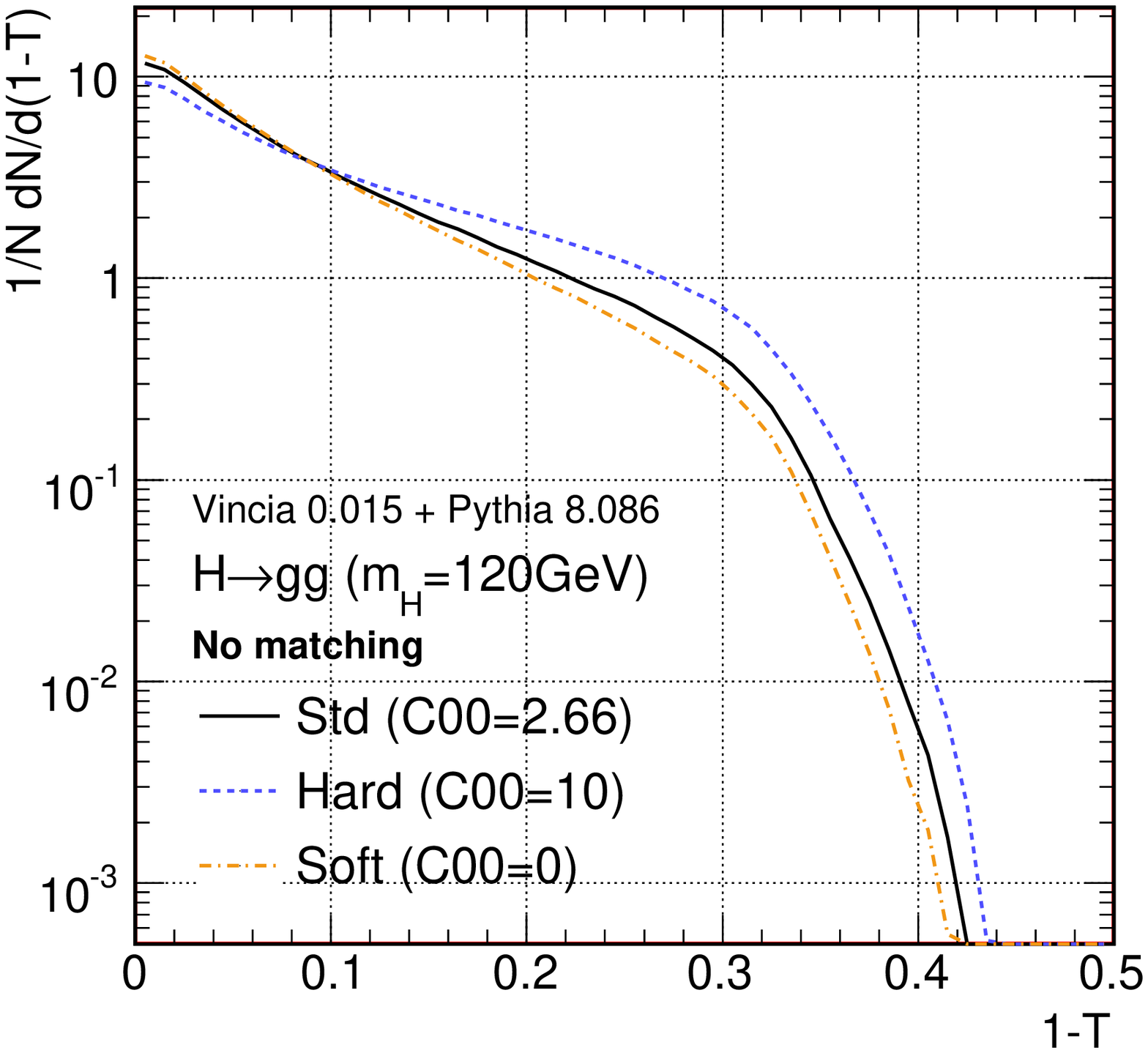}\hspace*{-5mm}&\hspace*{-5mm}
\includegraphics*[scale=0.43]{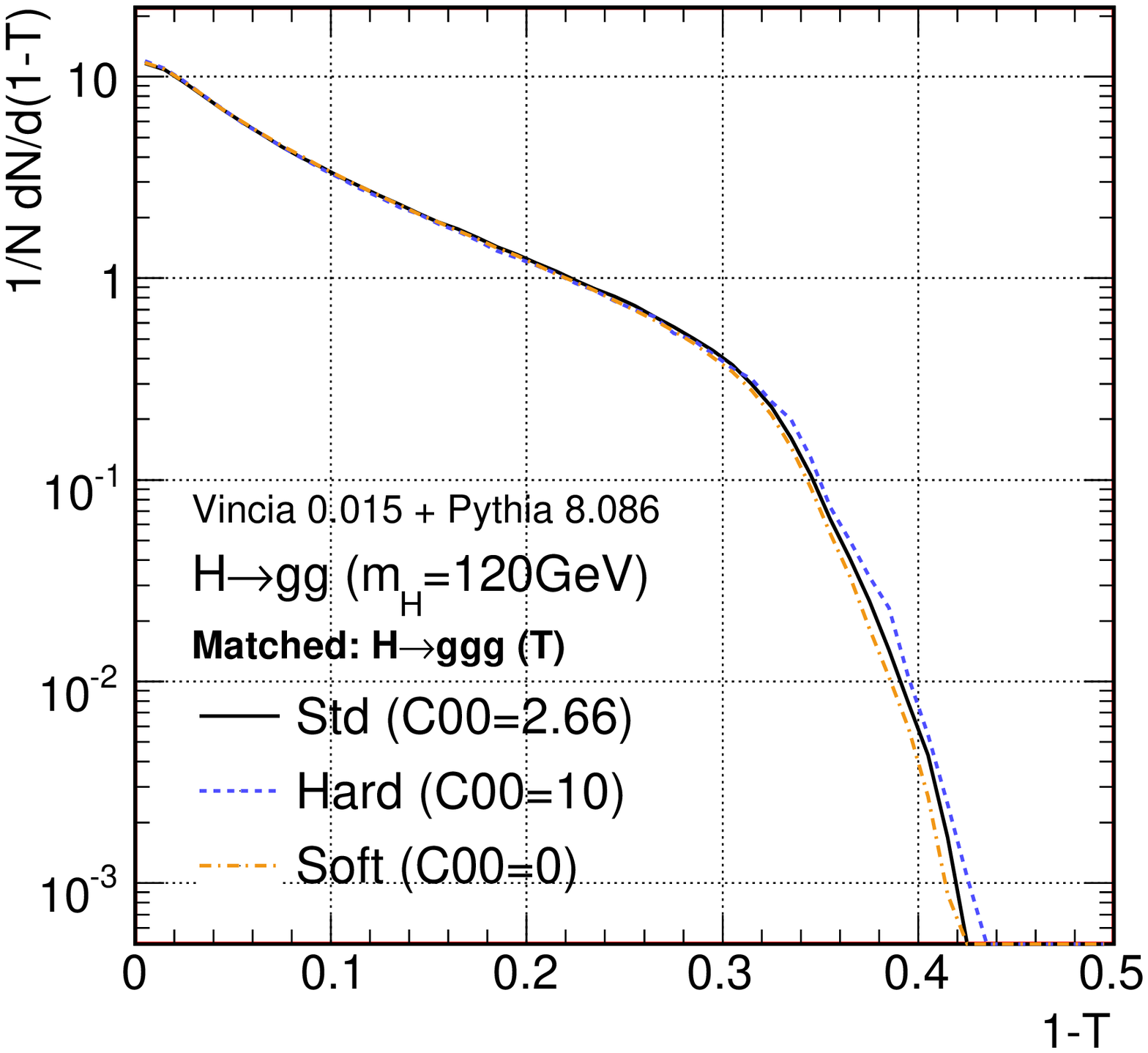}\hspace*{-1cm}
\end{tabular}\vspace{-7mm}
\caption{Differential Thrust ($1-T$) distribution. 
Labels are identical to those in fig.~\ref{fig:1stbranching}.
\label{fig:thrust}}
\end{center}
\end{figure}
The thrust is a more complicated observable, whose distribution is
shown in fig.~\ref{fig:thrust}, again with (right) and without (left)
matching. We see that, in the region accessible to 3-parton
configurations, $1-T<1/3$, the variation is indeed canceled, while
in the region of large $1-T$, accessible only to 4-parton and higher
configurations, some uncertainty remains (though with 2.5M events
generated per curve, small statistical 
fluctuations also become noticeable in the
sharply-dropping tail). We also note
that the behaviour at small $(1-T)$ is very sensitive to the
choice of $\mu_R$, a point to which we plan to return in the future. 
We have used $\mu_R=Q_\mrm{II}/2$ here. 

\section{Still Deeper? \label{sec:nll}}
In sect.~\ref{sec:matching}, we discussed how to match a leading-logarithmic
parton shower to tree-level calculations with an arbitrary preselected
number of resolved partons.  We also outlined how to perform a similar
matching at one loop.  In the previous section, we presented
a first implementation of these ideas.  While there is still a great
deal of work to be done in fleshing out and implementing our approach,
it is also interesting to peer ahead, and ask: how would one go further
in perturbation theory?  How could one further improve the accuracy
of parton-shower predictions?

One can presumably proceed to higher fixed orders, matching
to NNLO calculations by deriving generalizations of the equations
presented in sect.~\ref{sec:matching}.  We shall not examine
such matching in any greater detail.  Instead, let us explore
the possibility of resumming subleading logarithms, that is
including not only terms of $\Ord(\alpha_s^n \ln^{2n,2n-1} y)$
where $y$ is a large ratio of scales, but down to 
$\Ord(\alpha_s^{n+1}\ln^{2n,2n-1} y)$.

For this purpose, it is crucial to have a formalism that treats all
the leading-logarithmic singularities exactly point-by-point in
phase space.  This is true of the antenna-based formalism described
here (and would also be true of showering based on the Catani--Seymour
dipole formalism).  To set up a subleading-logarithmic shower,
we must consider corrections to the showering kernel
itself.  There are two kinds of corrections: virtual corrections,
and real-emission ones.  The former still correspond to a $2\rightarrow3$
branching process, but with the branching probability computed to
one order beyond leading in $\alpha_s$.  The real-emission corrections
correspond to a new branching process, $2\rightarrow4$ partons.  Such
a branching can occur with the basic Sudakov, of course, but only in
two branching steps.  Here, it will sometimes happen in one step.
Indeed, the kernel that will ultimately enter a modified Sudakov factor
will not be simply the $2\rightarrow4$ branching or antenna function,
but rather that function, with the iterated $2\rightarrow3$ contribution
subtracted out.  This excess represents the genuine {\it correlated\/}
$2\rightarrow4$ branching probability.  The required ingredients in
a dipole-antenna approach ---
one-loop corrections to the basic antenna, and the $2\rightarrow4$
tree-level antenna function --- are known from the development of
an NNLO fixed-order formalism~\cite{NNLOAntennae}.

We would further need a definition of the evolution variable that can be
evaluated on $n\rightarrow n+2$ branchings, and that regulates all
infrared divergences in them.  We would also need an appropriate
phase-space mapping for the following factorization,
\begin{equation}
\frac{\d\PS^{[i]}_{n+2}}{\d\PS_n} = \frac{\d\PS^{[i]}_{n-2}}{\d\PS_{n-2}}
  \frac{\d\PS^{[i]}_{4}}{\d\PS_{2}}~,\label{eq:dipolefactorizationNLL}
\end{equation}
which is now six- rather than three-dimensional.  With these
definitions and mappings, the NLL Sudakov would presumably take the form,
\begin{equation}
\begin{array}{rcl}\displaystyle
\Delta_\mrm{NLL}(t_{\mrm{n}},t_{\mrm{end}}) & = &\displaystyle 
 \Delta'_\mrm{LL}(t_{\mrm{n}},t_{\mrm{end}}) \ \times\\[3mm] & &
\hspace*{-1cm}
\displaystyle
\exp\Bigg[-\int_{t_{\mrm{in}}}^{t_\mrm{end}}\hspace*{-3mm} \d{t_{n+2}}
\hspace*{-3mm} \sum_{j\in \{n\to n+2\} } \hspace*{-1mm} \int 
\frac{\d\PS^{[j]}_{n+2}}{\d\PS_{n}} \ 
\delta(t_{n+2} - t^{[j]}(\{p\}_{n+2})) 
A^{[j]}_{2\to4}(\{p\}_{n}\!\!\to\!\{p\}_{n+2})\Bigg]~,\label{eq:SudakovNLL}
\end{array}
\end{equation}
where $\Delta'_\mrm{LL}$ includes the one-loop corrections to the LL
kernel. We will face the issue of fixing the finite terms in the LL
Sudakov factor upon matching to the NLL one, and a related issue of
maintaining the positivity of the resulting NLL corrections $A_{2\to4}$.
The requirement of maintained 
positivity is crucial to a probabilistic interpretation, 
such as the Markov chain.

Assuming these issues can be resolved in a satisfactory manner, the matching
prescriptions appear to generalize in a straightforward way.
At order $\alpha_s$, nothing much changes,
except for possible modifications to the LL kernels.
At relative order $\alpha_s^2$, the shower now
simply produces one extra term, corresponding to
a direct $2\to 4$ branching.  It contains the proper subleading
logarithms by construction, and accordingly the
tree-level matching equation will have the form (differentially in
$\d\PS_{X+2}$), 
\begin{equation}
\begin{array}{rcl}
\displaystyle
\Matching_{X+2}
&=&\displaystyle |M_{X+2}|^2
- \sum_i A^{[i]}_{2\to 3}(...)|M_{X+1}|^2
\Theta(t^{[i]}(\{p\}_{X+2})-t_{X+1}) \\[3mm]
& & \displaystyle
- \sum_j A^{[j]}_{2\to4}(...)|M_{X}|^2\Theta(t^{[j]}(\{p\}_{X+2})-t_{X})
~;~~~t<t_{\mrm{had}}~, \label{eq:mt2treeNLL}
\end{array}
\end{equation}
where $A^{[i]}_{2\to3}$ includes one-loop corrections and
$A^{[j]}_{2\to4}$ is the direct $2\to4$ kernel, as discussed above. 

At least from this perspective, the
inclusion of explicit $n\to n+2$ branchings poses no fundamental
problem.  In addition, inclusion of matching to one-loop and two-loop
matrix elements, which will both be modified by subtraction terms,
would open the way to event generation at NNLO.

\section{Conclusion and Outlook \label{sec:conclusion}}
We have presented a new general formalism for parton-shower
resummations.   The formalism allows us to explore both
the uncertainties inherent in the parton-shower predictions, 
and the reductions
in them possible by matching to fixed-order matrix elements.
We keep track of the ambiguities of the
shower approach away from the soft and collinear regions, allowing
us gain a systematic estimate of the associated uncertainties.  
The quantification of these uncertainties, as well of their reduction
by matching, is novel.

We have outlined a general approach for matching to fixed-order
matrix elements, based on a subtraction approach which generalizes
that of Frixione and Webber.
We also presented a specific algorithm based on antenna factorization
and dipole-antenna showers, generalizing that of Gustafson and L\"
onnblad. The formalism is simple and intuitive, but
is powerful enough to match fixed-order matrix elements at higher multiplicity
both at tree- and one-loop level.  In this respect, it
provides a generalization of both the CKKW and \Fw\ approaches.
The (arbitrary) choice of non-singular terms in the shower kernel
is explicitly canceled by the matching terms, which allows us to
quantify the degree to which
matching to a given order reduces the uncertainty inherent in parton-shower
predictions. 

We presented a generalization of the definition of
the hadronization cut-off that would make possible
a more universal modeling of non-perturbative physics, allowing more
meaningful comparisons of different parton-shower approaches, as
well as the improvement of fixed-order parton-level calculations without
reference to a specific hadronization model.

We have developed
a proof-of-concept level implementation for matching of
gluon showers in the decay process $H\to gg$ including both 
real and virtual corrections, in
the form of the \Vc\ code, and have presented
illustrative comparisons with and without matching for a
few benchmark distributions.

The next step will be to include quarks and perform a more
comprehensive study of both $H\to gg$ and $Z\to q\bar{q}$ 
fragmentation, exploring the properties of the 
\Vc\ algorithm and its relation to
existing approaches in greater detail.   We plan
to go into greater detail on various theoretical aspects
in a future paper \cite{VinciaPaper}. 
We believe that it should be straightforward
to automate matching for general lepton
collider and decay processes matrix elements, once the evolution variables are
generalized to be history-independent. 
The inclusion of initial-state radiation and matching
will be necessary to extend the approach to hadron collisions.
The formalism outlined here should be sufficiently general to make this
feasible.  Indeed, as we have discussed briefly, we believe it will 
be sufficiently general
to open a path to matching and showering deeper into the perturbative
regime, both in powers of $\alpha_s$ and orders of subleading logarithms.

\subsection*{Acknowledgments}
We thank Z.~Bern, R.~Frederix, A.~Gehrmann--De~Ridder, 
S.~Mrenna and T.~Sj\"ostrand for help and valuable comments.
W.~G.\ and P.~S.~ are supported by Fermi Research Alliance,
LLC, under Contract No.\ 
DE-AC02-07CH11359 with the United States Department of
Energy. D.~A.~K.\ is supported in part by the Agence Nationale de la
Recherche of France under 
grant ANR-05-BLAN-0073-01.  The {\em Service de Physique Th\'eorique} is a
laboratory of the {\em Direction des Sciences de la Mati\`ere} of the
{\em Commissariat \`a l'Energie Atomique} of France. 


\end{document}